\documentclass[prd,%
preprint,%
tightenlines,superscriptaddress,showpacs,%
nofootinbib,eqsecnum,amsfonts,amsmath]{revtex4} \usepackage{bm}
\usepackage{graphicx}

\begin{document}

\title{Hadamard regularization of the third post-Newtonian
gravitational wave generation of two point masses}

\date{\today}

\author{Luc Blanchet} \email{blanchet@iap.fr}
\affiliation{${\mathcal{G}}{\mathbb{R}}
\varepsilon{\mathbb{C}}{\mathcal{O}}$, Institut d'Astrophysique de
Paris (C.N.R.S.), 98$^{\text{bis}}$ boulevard Arago, 75014 Paris,
France}\affiliation{Yukawa Institute for Theoretical Physics, Kyoto
University, Kyoto 606-8502, Japan}

\author{Bala R. Iyer} \email{bri@rri.res.in} \affiliation{Raman Research
Institute, Bangalore 560080, India}

\begin{abstract}
Continuing previous work on the 3PN-accurate gravitational wave
generation from point particle binaries, we obtain the binary's 3PN
mass-type quadrupole and dipole moments for general (not necessarily
circular) orbits in harmonic coordinates. The final expressions are
given in terms of their ``core'' parts, resulting from the application
of the pure Hadamard-Schwartz (pHS) self-field regularization scheme,
and augmented by an ``ambiguous'' part. In the case of the 3PN
quadrupole we find three ambiguity parameters, $\xi$, $\kappa$ and
$\zeta$, but only one for the 3PN dipole, in the form of the
particular combination $\xi+\kappa$. Requiring that the dipole moment
agree with the center-of-mass position deduced from the 3PN equations
of motion in harmonic coordinates yields the relation
$\xi+\kappa=-9871/9240$. Our results will form the basis of the
complete calculation of the 3PN radiation field of compact binaries by
means of dimensional regularization.
\end{abstract}


\preprint{gr-qc/0409094}

\maketitle

\section{Introduction}\label{intro}

The present paper is the continuation of previous work
\cite{BIJ02}\,\footnote{Henceforth the Ref. \cite{BIJ02} will be
referred to as Paper I.} on the generation of gravitational waves by
inspiralling compact binaries, \textit{viz.} binary systems of neutron
stars or black holes whose orbit adiabatically spirals in by emission
of gravitational radiation. The adiabatic inspiral takes place right
before the final plunge and merger of the two compact objects, to
(presumably) form a single black hole which will settle down, after
emission of its quasi-normal modes, into a stationary
configuration. Inspiralling compact binaries will almost surely be
detected by large scale laser interferometric gravitational wave
observatories like VIRGO and LIGO. Recent estimates of the rate of
coalescences of neutron stars are very promising \cite{kalogera}.

It is by now well established (see \textit{e.g.}
Refs. \cite{3mn,CF94,TNaka94,P95,DIS98,DIS01,BCV03a,BCV03b,DIJS03})
that, in order to predict in a useful way the gravitational radiation
emitted by inspiralling compact binaries, general relativity must be
developed to high post-Newtonian (PN) order, probably up to the 3PN or
even the 3.5PN level.\,\footnote{As usual the $n$PN order refers to
the terms of order $1/c^{2n}$ in the waveform or energy flux,
relatively to the lowest-order (Newtonian) approximation described by
the Einstein quadrupole formula.} Correlatively it has been realized
that the crucial point for detecting and deciphering the gravitational
waves, is to accurately take into account the \textit{gravitational}
interaction between the compact objects --- responsible for the
binary's orbital dynamics and wave emission. Arguments within general
relativity \cite{D83houches} show that a good modelisation is by
\textit{point particles}, characterized only by mass parameters $m_1$
and $m_2$ (we neglect the intrinsic spins of the compact objects). It
makes sense to implement a model of point particles within the PN
approximation, provided that a process of regularization is used for
dealing with the infinite self-field of the point particles. The
regularization should hopefully be followed by a renormalization.

In paper I we adopted as self-field regularization the Hadamard
regularization \cite{Hadamard,Schwartz,Sellier}, augmented by a
prescription for adding a few arbitrary unknown ``\textit{ambiguity
parameters}'', accounting for the incompleteness of the Hadamard
regularization when evaluating certain divergent integrals occuring at
the 3PN order. We found that the 3PN mass quadrupole moment of
point-particle binaries is complete up to \textit{three} ambiguity
parameters, denoted $\xi$, $\kappa$ and $\zeta$, which would typically
be some rational fractions and could take, within Hadamard's
regularization, any numerical values. [The quadrupole moment is the
only one to be computed with full 3PN accuracy, thus it contains most
of the difficult non-linear integrals, and all the ambiguities
associated with the Hadamard regularization.] The gravitational-wave
flux, which is a crucial quantity to be predicted because it drives
the binary's orbital phase evolution, has then been found to be
complete, in the case of circular orbits, up to a single combination
of the latter ambiguity parameters, given by
$\theta=\xi+2\kappa+\zeta$. Of course the ambiguity parameters do not
affect the test mass limit of the result of paper I, which is found to
be in perfect agreement with the result of linear black-hole
perturbations in this limit \cite{TNaka94,Sasa94,TSasa94,TTS96}.

The parameters $\xi$, $\kappa$ and $\zeta$ represent the analogues,
for the case of the gravitational wave field (more precisely the mass
quadrupole moment), of similar parameters which were originally
introduced in the problem of the equations of motion of point particle
binaries at the 3PN order \cite{JaraS98,JaraS99,BF00,BFeom}. More
precisely, $\xi$ and $\kappa$ are the coefficients of some static
terms, independent of the particle's velocities but depending on their
accelerations, which can be viewed as some analogues of the ``static''
ambiguity constant $\omega_s$ in the 3PN ADM-Hamiltonian
\cite{JaraS98,JaraS99}, which is itself equivalent to the parameter
$\lambda$ entering the 3PN equations of motion in harmonic coordinates
\cite{BF00,BFeom}. On the other hand, $\zeta$ is the coefficient of a
particular velocity-dependent term in the mass quadrupole, and is
similar to the ``kinetic'' ambiguity $\omega_k$ in the Hamiltonian
\cite{JaraS98,JaraS99}, which has no counterpart in the equations of
motion since the velocity terms were unambiguously determined there
\cite{BF00,BFeom}. The work \cite{BF00,BFeom} used an improved version
of the Hadamard regularization called the extended-Hadamard
regularization, based on a theory of pseudo-functions and generalized
distributional derivatives, and defined in Refs. \cite{BFreg,BFregM}.

The ambiguity parameters in the binary's local dynamics (Hamiltonian and/or
equations of motion) have been resolved. For the kinetic ambiguity we have
$\omega_k=41/24$, which follows from the requirement of invariance under
global Poincar\'e transformations \cite{BF00,DJSpoinc}. On the other hand the
static ambiguity has been fixed using a powerful argument from dimensional
regularization, \textit{i.e.} computing the binary's dynamics in
$d=3+\varepsilon$ spatial dimensions and considering the limit where
$\varepsilon\rightarrow 0$, which led to $\omega_s=0$ \cite{DJSdim} or,
equivalently, to $\lambda=-1987/3080$ \cite{BDE04}. The same result has also
been achieved in Refs. \cite{IFA00,IFA01,itoh1,itoh2} by means of a surface
integral approach to the equations of motion of compact objects
(\textit{i.e.}, \textit{\`a la} Einstein-Infeld-Hoffmann), successfully
implemented at the 3PN order.

Summarizing, the 3PN equations of motion have been completed in
essentially two steps: the first one consists of using Hadamard's
regularization and permits the computation of most of the terms but a
few; the second step is to apply dimensional regularization in order
to fix the value of the few parameters left undetermined. In the
present situation it is not possible to compute the equations of
motion in the general $d$-dimensional case, but only in the limit
where $\varepsilon=d-3 \rightarrow 0$ \cite{DJSdim,BDE04}. In
Refs. \cite{DJSdim,BDE04} one computes the \textit{difference} between
the dimensional and Hadamard regularizations, and it is this
``difference'', specifically due to the existence of poles in $d$
dimensions (proportional to $1/\varepsilon$), which corresponds to the
ambiguities in Hadamard's regularization. Actually the latter
difference has to be defined with respect to a particular
Hadamard-type regularization of integrals called the ``\textit{pure
Hadamard-Schwartz}'' (pHS) regularization, following the terminology
and definition of Ref. \cite{BDE04}. The pHS regularization consists
of the standard notion of Hadamard's partie finie of divergent
integrals, together with a minimal treatment of the compact-support
(or ``contact'') terms, and the use of Schwartz distributional
derivatives \cite{Schwartz}. The result of dimensional regularization
is given by the sum of the pHS regularization and of the
``difference'' containing the poles proportional to $1/\varepsilon$.

In the present approach, for the 3PN wave generation we basically follow the
same two-step strategy as for the 3PN equations of motion, namely:
\begin{enumerate}
\item[(i)] To obtain the expression of the mass quadrupole moment at 3PN
order, as regularized by means of the pHS regularization;
\item[(ii)] To add to the pHS result the difference between the dimensional
regularization and the pHS one, which as we said above is due to the presence
of poles at the 3PN order.
\end{enumerate}
Imposing then that the result of dimensional regularization is
\textit{equivalent} to the result of (the pHS variant of) Hadamard's
regularization and augmented by appropriate ambiguity parameters, will then
uniquely determine the ambiguity parameters. A summary of the above
calculations leading to the following unique values for the ambiguity
parameters $\xi$, $\kappa$ and $\zeta$:
\begin{equation}\label{resdimreg}
\xi=-\frac{9871}{9240}\,,\qquad
\kappa=0\,,\qquad\zeta=-\frac{7}{33}\,,
\end{equation}
has been provided in Ref.~\cite{BDEI04}. The technical details of the above
calculations are now given in a series of three papers of which the present
one is the first. Indeed, the present paper is devoted to the calculation of
the pHS regularization of the 3PN quadrupole moment, \textit{item} (i) above.
The computation of the part associated with the difference between the
dimensional and pHS regularizations, \textit{cf. item} (ii), will be provided
in the third paper of this series \cite{BDEI04dr}. In the second paper
\cite{BDI04zeta} the value of $\zeta$ has been confirmed by a different
approach, based on the multipole moments of a boosted Schwarzschild solution.

It should be emphasized that the values (\ref{resdimreg}) represent the end
result of \textit{dimensional regularization}, obtained by means of the sum of
the steps (i) and (ii). However, we shall be able to obtain below [see
Eq.~(\ref{combamb}) and Section \ref{MD}] an independent confirmation, within
Hadamard's regularization, of the value of the particular combination of
parameters $\xi+\kappa$. Moreover, the fact that $\kappa=0$ has been checked
by a diagrammatic reasoning in Ref.~\cite{BDEI04dr}. Since as we said $\zeta$
has also been computed in \cite{BDI04zeta} by a different method, we see that
the present paper and the works \cite{BDI04zeta,BDEI04dr} altogether provide a
check, \textit{independent of dimensional regularization}, for all the
parameters (\ref{resdimreg}).

In our previous work (paper I), the 3PN mass quadrupole moment was regularized
by means of some ``hybrid'' Hadamard-type regularization, instead of the pHS
one, and the ambiguity parameters $\xi$, $\kappa$ and $\zeta$ were defined
with respect to that regularization. In the present paper, since we shall
perform a different computation, based on the specific pHS regularization, we
shall have to introduce some new ambiguity parameters. Since we do not want to
change the definition of $\xi$, $\kappa$ and $\zeta$, we shall perform some
numerical shifts of the values of $\xi$, $\kappa$ and $\zeta$, in order to
take into account the different reference points for their definition: hybrid
regularization in paper I, \textit{vs.} pHS regularization in the present
paper.

The present investigation will also extend and improve the analysis of paper I
in two important ways. First we shall use a better formulation of the
multipole moments at the 3PN order, in terms of a set of \textit{retarded}
elementary potentials, instead of the ``instantaneous'' versions of these
potentials as was done in paper I. The retarded potentials are the same as in
our computation of the equations of motion in harmonic coordinates
\cite{BFeom}; their use will appreciably simplify the present work. Secondly
we shall generalize paper I to the case of arbitrary orbits, not necessarily
circular. Circular orbits are in principle sufficient to describe the
inspiralling compact binaries, but the general non-circular case will be
mandatory when we want to obtain the values of $\xi$, $\kappa$ and $\zeta$
separately, and it is also important for a check of the over-all consistency
of our calculation.

Besides the 3PN mass quadrupole moment of point particle binaries we compute
also their 3PN mass \textit{dipole} moment. The mass dipole is interesting
because it is a conserved quantity (or it varies linearly with time), which is
already known from the conservative part of the binary's local 3PN equations
of motion in harmonic coordinates. Namely the dipole moment is nothing but the
integral of the \textit{center-of-mass} position associated with the
invariance of the equations of motion under the Poincar\'e group. It has been
computed from the binary's Lagrangian in harmonic coordinates at 3PN order in
Ref. \cite{DJSequiv,ABF01}. In fact what we shall do in the present paper is
to \textit{impose} the equivalence between the 3PN dipole moment and the 3PN
center-of-mass vector position, and we shall prove that this requirement fixes
uniquely \textit{one}, but only one, combination of the ambiguity parameters,
\textit{viz.}
\begin{equation}\label{combamb}
\xi+\kappa = -\frac{9871}{9240}\,,
\end{equation}
working \textit{solely within Hadamard regularization}. This result is
perfectly consistent with the complete result provided by dimensional
regularization \cite{BDEI04} and recalled in Eq.~(\ref{resdimreg}). We view
this agreement as an important check of the correctness of both the present
calculation and the one of Ref. \cite{BDEI04}.

The paper is organized as follows. In Section \ref{multdecomp} we
review our definitions of the multipole moments of an isolated
extended source in the PN approximation, both time-vaying moments
(having $\ell\geq 2$) and static ones ($\ell=0,1$). In Section
\ref{3PNmoment}, we give the explicit expression of the mass-type
moments in terms of a set of retarded elementary potentials up to 3PN
order. The pure-Hadamard-Schwartz (pHS) regularization scheme is then
reviewed in Section \ref{Had}, where we comment on the various types
of terms encountered in the calculation, and we detail our practical
way to perform the partie finie of three dimensional non-compact
support spatial integrals. Our final results for both the 3PN
quadrupole and dipole moments, and our derivation of
Eq. (\ref{combamb}), are presented in Section \ref{pointpart}. The
formula for the 3PN quadrupole moment in a general frame turned out to
be too long to be published, therefore we choose to present it in the
frame of the center of mass (but for general orbits): see
Eqs. (\ref{Iijfinal})--(\ref{ABC}) below.

\section{Multipole decomposition of the exterior field}\label{multdecomp}

In this Section we provide an account of the relevant notion of
multipole moments of a general isolated gravitational wave source. By
definition the moments parametrize the linearized approximation in a
post-Minkowskian expansion scheme for the gravitational field in the
external (vacuum) domain of the source. Their explicit expressions in
terms of the source's physical parameters (matter stress-energy tensor
$T^{\mu\nu}$) have been found in the case of a \textit{post-Newtonian}
source by using a variant of the theory of matched asymptotic
expansions \cite{B95,B98mult,PB02}. The matching relates the exterior
field of the source, as obtained from a multipolar-post-Minkowskian
expansion of the external field \cite{BD86}, to the inner field of the
post-Newtonian extended source, as iterated in the standard PN way.

\subsection{External solution of the field equations}\label{multsol}

The Einstein field equations are cast into ``relaxed'' form by means
of the condition of harmonic (or De Donder) coordinates. Denoting the
fundamental gravitational field variable by $h^{\mu\nu}\equiv
\sqrt{-g}\,g^{\mu\nu}-\eta^{\mu\nu}$,\,\footnote{Here, $g^{\mu\nu}$
denotes the inverse of the usual covariant metric $g_{\mu\nu}$; $g$ is
the determinant of $g_{\mu\nu}$: $g=\mathrm{det}(g_{\mu\nu})$; and
$\eta^{\mu\nu}$ is an auxiliary Minkowskian metric in Minkowskian
coordinates: $\eta^{\mu\nu}=\mathrm{diag}(-1,1,1,1)$.} this means that
\begin{equation}\label{harm}
\partial_\nu h^{\mu\nu}=0\,.
\end{equation}
Under the harmonic coordinate conditions the field equations take the
form of non-linear wave equations
\begin{equation}\label{EE}
\Box h^{\mu\nu}=\frac{16\pi G}{c^4}\tau^{\mu\nu}\,,
\end{equation}
in which $\Box\equiv\eta^{\rho\sigma}\partial_{\rho\sigma}$ denotes
the standard flat space-time d'Alembertian operator. The
right-hand-side of Eq. (\ref{EE}) is made of the total (matter plus
gravitation) pseudo stress-energy tensor in harmonic coordinates given
by
\begin{equation}\label{tau}
\tau^{\mu\nu} = \vert g\vert T^{\mu\nu}+\frac{c^4}{16\pi
G}\Lambda^{\mu\nu}\bigl[h,\partial h,\partial^2 h\bigr]\,,
\end{equation}
where $T^{\mu\nu}$ is the stress-energy tensor of the matter fields,
and $\Lambda^{\mu\nu}$ represents the gravitational source term which
is given by a complicated non-linear, quadratic at least, functional
of $h^{\rho\sigma}$ and its first and second space-time
derivatives. Equation (2.3) in paper I gives the explicit expression
of $\Lambda^{\mu\nu}$. In the following we shall assume that the
support of $T^{\mu\nu}$ is spatially compact. In our formalism the
conservation of the pseudo tensor,
\begin{equation}\label{dtau}
\partial_\nu\tau^{\mu\nu} = 0\,,
\end{equation}
is the consequence of the harmonic-coordinate condition (\ref{harm}).

Let the calligraphic letter $\mathcal{M}$ denote the operation of
taking the multipole expansion, so that $\mathcal{M}(h^{\mu\nu})$
represents the multipole expansion of the external gravitational field
--- a solution of the \textit{vacuum} field equations valid outside
the compact support of the matter tensor $T^{\mu\nu}$. Similarly
${\cal M}(\Lambda^{\mu\nu})$ denotes the multipole expansion of the
gravitational source term, and is obtained from insertion into
$\Lambda^{\mu\nu}$ of the multipole expansions of $h$ and its
space-time derivatives. Note that ${\cal M}(T^{\mu\nu})=0$ since
$T^{\mu\nu}$ has a compact support. We want to compute the multipole
moments of an extended \textit{post-Newtonian} source (one for which
the PN approximation is physically meaningful). To this end we first
consider the following quantity:
\begin{equation}\label{Delta}
\Delta^{\mu\nu}\equiv h^{\mu\nu}-\mathop{\mathrm{FP}}_{B=0}\,
\Box^{-1}_\mathcal{R} \Bigl[ \widetilde r^B {\cal
M}(\Lambda^{\mu\nu})\Bigr]\,,
\end{equation}
which is made of the difference between $h^{\mu\nu}$, the solution of
the field equations (\ref{EE}) valid everywhere inside and outside the
source, and a particular object obtained from the \textit{retarded}
($\mathcal{R}$) integral of the \textit{multipole} ($\mathcal{M}$)
expansion of the gravitational source term $\Lambda^{\mu\nu}$. Here
the retarded integral means the usual flat space-time expression
\begin{equation}\label{ret}
\Box^{-1}_\mathcal{R}f(\mathbf{x},t)=-\frac{1}{4\pi}
\int\frac{d^3\mathbf{x}'}
{\vert\mathbf{x}-\mathbf{x}'\vert}f\biggl(\mathbf{x}',
t-\frac{\vert\mathbf{x}-\mathbf{x}'\vert}{c}\biggr)\,.
\end{equation}
Since Eq. (\ref{ret}) extends up over the whole space,
$\mathbf{x}'\in\mathbb{R}^3$, including the region inside the source
where the multipole-moment expansion is not valid (it diverges at the
spatial origin $\vert\mathbf{x}'\vert\rightarrow 0$ located inside the
source), one is not allowed to use directly the retarded integral as
it stands. This is the reason for the introduction in the second term
of Eq. (\ref{Delta}) of a particular regularization process defined by
the \textit{finite part} when a complex number $B$ tends to zero (this
operation is abbreviated as $\mathop{\mathrm{FP}}_{B=0}$), and
involving the regularization factor
\begin{equation}\label{rtilde}
\widetilde{r}^B
\equiv\vert\widetilde{\mathbf{x}}\vert^B\equiv\left(\frac{r}{r_0}\right)^B\,,
\end{equation}
which is to be inserted in front of the multipolar-expanded source
term of the retarded integral. Here $r_0$ denotes an arbitrary
constant scale having the dimension of a length. Since the divergence
of the retarded integral is at the origin of the coordinates,
$\vert\mathbf{x}'\vert\rightarrow 0$ in (\ref{ret}), the constant
$r_0$ plays the role of an ultra-violet (UV) cut-off in the second
term of Eq. (\ref{Delta}). However we shall see that in the expression
of the multipole moments themselves, given by Eqs. (\ref{HL}) or
(\ref{FL}) below, the \textit{same constant} $r_0$ will appear to
represent in fact an \textit{infra-red} (IR) cut-off.

From Eq. (\ref{Delta}), and noticing that the second term is already
in the form of a multipole expansion, we can write the complete
multipole decomposition of the external field as
\begin{equation}\label{multh}
\mathcal{M}(h^{\mu\nu}) = \mathcal{M}(\Delta^{\mu\nu}) +
\mathop{\mathrm{FP}}_{B=0}\, \Box^{-1}_\mathcal{R} \Bigl[ \widetilde
r^B {\cal M}(\Lambda^{\mu\nu})\Bigr]\,.
\end{equation}
This is a solution of the vacuum Einstein field equations,
\begin{subequations}\label{EEvacuum}\begin{eqnarray}
\Box \mathcal{M}(h^{\mu\nu})&=&\mathcal{M}(\Lambda^{\mu\nu})\,,\\
\partial_\nu\mathcal{M}(h^{\mu\nu})&=&0\,,\label{harmmult}
\end{eqnarray}\end{subequations}
valid in the exterior of the source where ${\cal
M}(T^{\mu\nu})=0$. The first term in Eq. (\ref{multh}) is given by an
homogeneous solution of the wave equation: $\Box
\mathcal{M}(\Delta^{\mu\nu})=0$. The second term in (\ref{multh})
represents a particular, inhomogeneous, solution, which arises because
of the nonlinearities in the external gravitational field, and can be
computed by means of the multipolar-post-Minkowskian algorithm of
Ref. \cite{BD86}. In the present paper we shall not consider the
second term in (\ref{multh}) because its contribution, encompassing
many non-linear effects, has already been computed in \cite{B98tail}
up to 3.5PN order for compact binaries. We shall define our source
multipole moments from the contribution ${\cal M}(\Delta^{\mu\nu})$,
which can be viewed in fact as the ``linearized'' part of the
multipolar decomposition (\ref{multh}).

It has been proved in Refs. \cite{B95,B98mult,PB02} that: (i) the
multipole expansion $\mathcal{M}(\Delta^{\mu\nu})$ of the quantity
defined by Eq. (\ref{Delta}) can be computed using the standard
formulas (given for instance in Refs. \cite{BD89} and \cite{DI91b})
for the multipole expansion outside a \textit{compact support} source;
(ii) the multipole moments admit a very simple expression in the case
where the matter source is slowly-moving (existence of a small PN
parameter $\varepsilon\sim v/c$). The result we find reads
as\,\footnote{The notation is: $L\equiv i_1\cdots i_\ell$ for a
multi-index composed of $\ell$ multipolar indices $i_1, \cdots,
i_\ell$; $\partial_L\equiv\partial_{i_1}\cdots \partial_{i_\ell}$ for
the product of $\ell$ partial derivatives
$\partial_i=\partial/\partial x^i$; similarly $x_L\equiv x_{i_1}\cdots
x_{i_\ell}$ for the product of $\ell$ spatial vectors $x^i\equiv
x_i$. In the case of summed-up (dummy) multi-indices $L$, we do not
write the $\ell$ summation symbols, from 1 to 3, over their indices.}
\begin{equation}\label{Mdelta}
\mathcal{M}(\Delta^{\mu\nu}) = - \frac{4G}{c^4}
\sum^{+\infty}_{\ell=0}\frac{(-)^\ell}{\ell!}\, \partial_L \left[
\frac{1}{r} \mathcal{H}^{\mu\nu}_L \Bigl(t-\frac{r}{c}\Bigr)
\right]\,,
\end{equation}
where the time-dependent functionals $\mathcal{H}^{\mu\nu}_L$ so
introduced, which depend on the retarded time $u=t-r/c$, are
explicitly given by\,\footnote{With a slight abuse of notation the
generic source-point on which one integrates in (\ref{HL}) is denoted
by $\mathbf{x}$ which is not the same as the field-point appearing in
the R.H.S. of Eq. (\ref{Mdelta}).}
\begin{equation}\label{HL}
\mathcal{H}^{\mu\nu}_L (u) = \mathop{\mathrm{FP}}_{B=0} \int d^3
\mathbf{x}\, \vert\widetilde{\mathbf{x}}\vert^B x_L \,
\overline{\tau}^{\mu\nu} (\mathbf{x}, u)\,.
\end{equation}

The integrand of the multipolar functional (\ref{HL}) involves the
\textit{post-Newtonian} expansion of the total pseudo stress-energy tensor
given by (\ref{tau}), where the formal operation of taking the PN expansion is
denoted by means of an overbar, \textit{i.e.} $\overline{\tau}^{\mu\nu}\equiv
\mathrm{PN}\left[\tau^{\mu\nu}\right]$. This is the crucial point on which we
recognize that the expression (\ref{HL}) is valid only for extended PN
sources, whose compact support extends well within their own near zone (see
\cite{B98mult} for details). The other important feature of Eq. (\ref{HL}) is
the presence of the finite part operation when $B\rightarrow 0$, with
regularization factor given by (\ref{rtilde}). The role of the finite part is
to deal with the IR divergences initially introduced in the multipole moments
by the fact that the PN-expanded integrand of the multipole moments diverges
at spatial infinity (when $r\rightarrow +\infty$). By contrast, we recall that
the finite part in the second term of (\ref{Delta}) was to take care of the UV
divergences when $r\rightarrow 0$. The fact that the same finite part
operation appears to be either IR or UV depending on the formula is made
possible by the properties of the complex analytic continuation (with respect
to $B\in\mathbb{C}$). We also mention the fact that the two terms in Eq.
(\ref{multh}) depend separately on the length scale $r_0$, but that this
dependence is in fact fictitious because the $r_0$'s can be shown to cancel
out. [To prove this the best way is to formally differentiate the R.H.S. of
Eq. (\ref{multh}) with respect to $r_0$.]

\subsection{The STF source multipole moments}\label{STF}

In the present approach it is convenient to work with the equivalent
of the multipole expansion (\ref{Mdelta})--(\ref{HL}) but written in
symmetric and trace-free (STF) guise. We present only the results. For
the multipole expansion we have
\begin{equation}\label{multDeltaF}
\mathcal{M}(\Delta^{\mu\nu}) = - \frac{4G}{c^4}
\sum^{+\infty}_{\ell=0}\frac{(-)^\ell}{\ell!}\, \partial_L \left[
\frac{1}{r} \mathcal{F}^{\mu\nu}_L \Bigl(t-\frac{r}{c}\Bigr)
\right]\,,
\end{equation} 
where the multipole moment functionals $\mathcal{F}^{\mu\nu}_L(u)$ are
now STF with respect to their $\ell$ indices $L=i_1\cdots i_\ell$. The
$\mathcal{F}^{\mu\nu}_L$'s differ from their counterparts
$\mathcal{H}^{\mu\nu}_L$ parametrizing the non-STF multipole
decomposition (\ref{Mdelta}). They are given by\,\footnote{We denote
the symmetric-trace-free (STF) projection by means of a hat,
$\hat{x}_L\equiv\mathrm{STF}(x_{i_1}\cdots x_{i_\ell})$, or sometimes
by means of brackets $\langle\rangle$ surrounding the indices,
$\hat{x}_L\equiv x_{\langle L\rangle}$.}
\begin{equation}\label{FL}
\mathcal{F}^{\mu\nu}_L (u) = \mathop{\mathrm{FP}}_{B=0} \int d^3
\mathbf{x}\, \vert\widetilde{\mathbf{x}}\vert^B \hat{x}_L \,
\int^1_{-1} dz\, \delta_l(z) \,\overline{\tau}^{\mu\nu} (\mathbf{y},
u+z\vert \mathbf{x}\vert /c)\,.
\end{equation} 
Equation (\ref{FL}) involves an extra integration, with respect to its
non-STF counterpart (\ref{HL}), over the variable $z$, and with
associated ``weighting'' function $\delta_\ell (z)$ given by
\begin{subequations}\label{deltal}\begin{eqnarray}
\delta_\ell (z) &\equiv& \frac{(2\ell+1)!!}{2^{\ell+1} \ell!}
\,(1-z^2)^\ell\,, \\ \int^1_{-1} dz\,\delta_\ell
(z)&=&1\,,\\\lim_{\ell \rightarrow +\infty}\delta_\ell (z)&=&\delta
(z)\,.
\end{eqnarray}\end{subequations}
Here $\delta(z)$ is Dirac's one-dimensional delta-function.

To obtain the \textit{source multipole moments}, we decompose the
function $\mathcal{F}^{\mu\nu}_L$, which is already STF in its $\ell$
indices composing $L$, into STF irreducible pieces with respect to all
its spatial indices, including those coming from the space-time
indices $\mu\nu=\{00,0i,ij\}$. The appropriate decompositions
read\,\footnote{We denote by $\varepsilon_{ijk}$ the usual Levi-Civita
anti-symmetric symbol such that $\varepsilon_{123}=+1$.}
\begin{subequations}\label{STFdecomp}
\begin{eqnarray}
\mathcal{F}^{00}_L &=& R_L\,, \\ \mathcal{F}^{0i}_L &=& {}^{(+)}T_{iL}
+ \varepsilon_{ai<i_l} {}^{(0)}T_{L-1>a} +\delta_{i<i_l}
{}^{(-)}T_{L-1>}\,,\\ {\cal F}^{ij}_L &=& {}^{(+2)}U_{ijL} +
\displaystyle{\mathop{\mathrm{STF}}_L}\,
\displaystyle{\mathop{\mathrm{STF}}_{ij}}\,
\Bigl[\varepsilon_{aii_l}{}^{(+1)}U_{ajL-1} +\delta_{ii_l}
{}^{(0)}U_{jL-1}\nonumber \\ && +\delta_{ii_l} \varepsilon_{aji_{l-1}}
{}^{(-1)}U_{aL-2} +\delta_{ii_l} \delta_{ji_{l-1}} {}^{(-2)}U_{L-2}
\Bigr] + \delta_{ij} V_L\,.
\end{eqnarray}\end{subequations}
We have introduced ten STF tensors $R_L$, ${}^{(+)}T_{L+1}$, $\cdots$,
${}^{(-2)}U_{L-2}$, $V_L$, equivalent to the ten components of the
original tensor $\mathcal{F}^{\mu\nu}_L$. Because of the
harmonic-gauge condition (\ref{harmmult}), only six of these tensors
are independent, and we are led to a set of six source multipole
moments, denoted $\{I_L,\,J_L,\,W_L,\,X_L,\,Y_L,\,Z_L\}$. These
moments are defined in such a way \cite{B98mult} that the four last
ones, $\{W_L,\,X_L,\,Y_L,\,Z_L\}$, parametrize a mere linearized gauge
transformation of the ``linearized'' part of the multipolar metric,
and consequently do not play a very important role. In practice the
moments $\{W_L,\,\cdots,\,Z_L\}$ appear only at high PN order, where
they can be typically computed with Newtonian precision, so they do
not pose computational problems. They have already been taken care of
in paper I.

The ``main'' source multipole moments are the mass-type moment $I_L$
and the current-type one $J_L$. In Section \ref{3PNmoment} we shall
concentrate our attention on the mass moment $I_L$ with full 3PN
accuracy. Having in hands the STF-irreducible decompositions
(\ref{STFdecomp}), we obtain in the generic case where $\ell\geq 2$
(\textit{i.e.} for non-conserved, arbitrary \textit{time-varying},
moments):
\begin{subequations}\label{ILJL0}\begin{eqnarray}
I_L &=& \frac{1}{c^2} \bigl(R_L + 3V_L\bigr) - \frac{4}{c^3(\ell
+1)}\,{}^{(-)} \dot{T}_L+ \frac{2}{c^4(\ell +1)(\ell+2)}\,{}^{(-2)}
\ddot{U}_L \,,\\ J_L &=& -\frac{\ell+1}{\ell c}\,{} ^{(0)}T_L
+\frac{1}{2\ell c^2}\,{}^{(-1)} \dot{U}_L\,,
\end{eqnarray}\end{subequations} 
where time derivatives are indicated by dots. To express the results
(\ref{ILJL0}) in the best way, we introduce the following definitions,
\begin{subequations}\label{Sigma}\begin{eqnarray}
\Sigma &\equiv& \frac{\overline{\tau}^{00}
+\overline{\tau}^{ii}}{c^2}~~\hbox{(where $\overline{\tau}^{ii}\equiv
\delta_{ij}\overline{\tau}^{ij}$)}\,,\\ \Sigma_i &\equiv&
\frac{\overline{\tau}^{0i}}{c}\,,\\ \Sigma_{ij} &\equiv&
\overline{\tau}^{ij}\,.\end{eqnarray}\end{subequations} 
For simplicity's sake we omit the overbar of the $\Sigma_{\mu\nu}$'s
to indicate the post-Newtonian expansion, but we do not forget that
these quantities are given by, and should be treated as,
\textit{formal PN-expanded} expressions. The STF source moments, for
multipolarities $\ell\geq 2$, are then given by \cite{B98mult}
\begin{subequations}\label{ILJL}\begin{eqnarray}
I_L(u)&=& \mathop{\mathrm{FP}}_{B=0}\,\int
d^3\mathbf{x}\,\vert\widetilde{\mathbf{x}}\vert^B \int^1_{-1}
dz\left\{ \delta_\ell\,\hat{x}_L\,\Sigma
-\frac{4(2\ell+1)}{c^2(\ell+1)(2\ell+3)} \,\delta_{\ell+1}
\,\hat{x}_{iL} \,\dot{\Sigma}_i\right.\nonumber\\ &&\qquad\quad
\left. +\frac{2(2\ell+1)}{c^4(\ell+1)(\ell+2)(2\ell+5)}
\,\delta_{\ell+2}\,\hat{x}_{ijL}\ddot{\Sigma}_{ij}\right\}
(\mathbf{x},u+z \vert{\mathbf{x}}\vert/c)\,,\label{IL}\\J_L(u)&=&
\mathop{\mathrm{FP}}_{B=0}\,\varepsilon_{ab<i_\ell} \int d^3
\mathbf{x}\,\vert\widetilde{\mathbf{x}}\vert^B \int^1_{-1} dz\biggl\{
\delta_\ell\,\hat{x}_{L-1>a} \,\Sigma_b \nonumber\\ &&\qquad\quad
-\frac{2\ell+1}{c^2(\ell+2)(2\ell+3)}
\,\delta_{\ell+1}\,\hat{x}_{L-1>ac} \,\dot{\Sigma}_{bc}\biggr\}
(\mathbf{x},u+z \vert\mathbf{x}\vert/c)\,,\label{JL}
\end{eqnarray}\end{subequations}
where the $\Sigma_{\mu\nu}$'s are evaluated at the position
$\mathbf{x}$ and at time $u+z \vert{\mathbf{x}}\vert/c$. In the
limiting case of \textit{linearized} gravity, we can replace
$\overline{\tau}^{\mu\nu}$ by the compact-support matter tensor
$T^{\mu\nu}$ and ignore the finite part procedure
($\mathop{\mathrm{FP}}_{B=0}$), so we recover the linearized-gravity
expressions obtained in \cite{DI91b}. Let us emphasize that
Eqs. (\ref{ILJL}) are ``\textit{exact}'', in the sense that they are
formally valid up to any PN order. In practice, the PN-expanded
moments (\ref{ILJL}) are to be computed by means of the infinite
post-Newtonian series
\begin{subequations}\label{intdeltal}\begin{eqnarray}
\int^1_{-1} dz~ \delta_\ell(z) \,\Sigma(\mathbf{x},u+z
\vert\mathbf{x}\vert/c) &=& \sum_{k=0}^{+\infty}\,\alpha_{k,\ell}
\,\left(\frac{\vert\mathbf{x}\vert}{c}\frac{\partial}{\partial
u}\right)^{2k} \!\Sigma(\mathbf{x},u)\,,\\ \alpha_{k,\ell} &\equiv&
\frac{(2\ell+1)!!}{(2k)!!(2\ell+2k+1)!!}\,.
\end{eqnarray}\end{subequations}
In a separate work \cite{BDI04zeta} we shall derive some alternative
expressions of the PN moments (\ref{ILJL}) in the form of integrals depending
only on the boundary at infinity (\textit{i.e.}
$\vert\mathbf{x}\vert\rightarrow +\infty$, $u=\mathrm{const}$).

\subsection{The conserved monopole and dipole moments}\label{monodipo}

In the case of the non-radiative moments, \textit{i.e.} the mass
monopole $M$ ($\ell=0$) and the mass and current dipoles $M_i$ and
$S_i$ ($\ell=1$), things are a little bit more involved than what is
given by Eqs. (\ref{ILJL}). The monopole and dipoles are conserved by
virtue of the source's equation of motion, Eq. (\ref{dtau}), namely
\begin{subequations}\label{conslaw}\begin{eqnarray}
\dot{M} &=& 0\,,\\ \ddot{M}_i &=& 0\,,\\ \dot{S}_i &=& 0\,.
\end{eqnarray}\end{subequations}
In particular $M$ denotes the ADM mass of the source. As shown in
\cite{B98mult} the conserved monopole and dipoles can be written into
the form
\begin{subequations}\label{Icons}\begin{eqnarray}
M &=& I + \delta I\,,\\ M_i &=& I_i + \delta I_i\,,\\ S_i &=& J_i +
\delta J_i\,,
\end{eqnarray}\end{subequations}
where the first pieces $I$, $I_i$ and $J_i$ are \textit{defined} by
the same formulas as Eqs. (\ref{ILJL}) but in which we set either
$\ell=0$ or $\ell=1$, and where the extra pieces follow from
Eqs. (5.6) in \cite{B98mult}, together with (5.4) and (4.5) there, and
are explicitly given by
\begin{subequations}\label{DeltaIcons}\begin{eqnarray}
\delta I &=& \mathop{\mathrm{FP}}_{B=0}\,B\int
d^3\mathbf{x}\,\vert\widetilde{\mathbf{x}}\vert^B\,\frac{x_a}
{\vert\mathbf{x}\vert^2}\int^1_{-1}
dz\left\{-\delta_0\,\Sigma_a^{(-1)}+\frac{1}{c^2}\,\delta_1
\,x_b\,\Sigma_{ab}\right\}\,,\\ \delta I_i &=&
\mathop{\mathrm{FP}}_{B=0}\,B\int
d^3\mathbf{x}\,\vert\widetilde{\mathbf{x}}\vert^B\,\frac{x_a}
{\vert\mathbf{x}\vert^2}\int^1_{-1}
dz\left\{-\delta_1\,x_i\,\Sigma_a^{(-1)}-\delta_0\,\Sigma_{ia}^{(-2)}
+\frac{1}{c^2}\,\delta_2\,\hat{x}_{ib}\,\Sigma_{ab}\right\}\,,~~
\label{DeltaIicons}\\ \delta J_i &=& \mathop{\mathrm{FP}}_{B=0}\,B\int
d^3\mathbf{x}\,\vert\widetilde{\mathbf{x}}\vert^B\,
\varepsilon_{iab}\,\frac{x_{bc}} {\vert\mathbf{x}\vert^2}\int^1_{-1}
dz\,\delta_1\,\Sigma_{ac}^{(-1)}\,.
\end{eqnarray}\end{subequations}
Time anti-derivatives are denoted by superscripts $(-n)$; $\delta_0$
and $\delta_1$ refer to the function given by (\ref{deltal}); like in
Eqs. (\ref{ILJL}) the integrands are evaluated at point $\mathbf{x}$
and at time $u+z \vert{\mathbf{x}}\vert/c$. The quantities $\delta I$,
$\delta I_i$ and $\delta J_i$ are precisely such that the ``total''
moments $M$, $M_i$ and $S_i$ obey the conservation laws
(\ref{conslaw}) as a consequence of the matter equations of motion
(see Ref. \cite{B98mult} for further discussion).

The chief feature of the expressions (\ref{DeltaIcons}) is that they
involve an explicit factor $B$ in front, and therefore they depend
only on the behavior of the integrand when
$\vert\mathbf{x}\vert\rightarrow +\infty$, since they are zero unless
the integral develops a pole $\sim 1/B$ due to the behavior of the
integrand near the boundary at infinity. We shall give more details on
the way we compute such integrals ``at infinity'' in Section
\ref{infbound}.

\section{The mass multipole moments at the 3PN order}\label{3PNmoment}

In this Section we derive the mass-type source multipole moment $I_L$
(for arbitrary $\ell\geq 2$) at the 3PN approximation, for general
extended PN sources. From Eqs. (\ref{ILJL}) we see that one must
obtain $\Sigma$ with full 3PN accuracy, but $\Sigma_i$ at the 2PN
order only, and $\Sigma_{ij}$ at 1PN order. For this purpose, we make
explicit the components of the $\Sigma_{\mu\nu}$'s, defined by
Eqs. (\ref{Sigma}), in terms of a certain set of retarded
``elementary'' potentials:
$V,\,V_i,\,\hat{W}_{ij},\,\hat{X},\,\hat{R}_{i},\,\hat{Z}_{ij}$,
solutions of appropriate iterated d'Alembertian equations. Although
devoid of any direct physical meaning, these potentials have proved to
constitute some very useful ``building blocks'' for practical PN
calculations on gravitational-wave generation (paper I), as well as in
the problem of equations of motion \cite{BFP98,BFeom}. In paper I we
systematically expanded all the retardations in $V,\,V_i,\,\cdots$ and
introduced some associated ``Poisson-like'' potentials
$U,\,U_i,\,\cdots$. Here we shall come back to the same retarded-like
potentials $V,\,V_i,\,\hat{W}_{ij},\,\cdots,\,\hat{Z}_{ij}$ as in the
equations of motion; of course we are motivated by the fact that they
have already been computed in \cite{BFeom}. So we shall redo entirely
the computation of paper I, using different elementary potentials and
also more systematic Mathematica programs, and in the case of general
orbits. Our results will match perfectly with those of paper I.

Let us denote the ``matter'' parts in the total density contributions
(\ref{Sigma}) by
\begin{subequations}\label{sigma}\begin{eqnarray}
\sigma &\equiv& \frac{T^{00} +T^{ii}}{c^2}\,;~~ T^{ii}\equiv
\delta_{ij}T^{ij}\,,\\ \sigma_i &\equiv& \frac{T^{0i}}{c}\,,\\
\sigma_{ij} &\equiv& T^{ij}\,.\end{eqnarray}\end{subequations} Our
chosen definitions for the elementary retarded-type potentials, which
involve non-linear couplings appropriate to 3PN order, are
\allowdisplaybreaks{\begin{subequations}\label{potentials}\begin{eqnarray}
V &=& \Box^{-1}_\mathcal{R} \bigl[-4\pi G\sigma \bigr]\,,\label{V}\\
V_i &=& \Box^{-1}_\mathcal{R}\bigl[-4\pi G
\sigma_i\bigr]\,,\label{Vi}\\ \hat{W}_{ij} &=&
\Box^{-1}_\mathcal{R}\Bigl[-4 \pi G \bigl(\sigma_{ij} - \delta_{ij}
\sigma_{kk}\bigr) - \partial_i V \partial_j V\Bigr]\,,\label{Wij}\\
\hat{X} &=& \Box^{-1}_\mathcal{R}\Bigl[ - 4\pi G \sigma_{ii}V +
\hat{W}_{ij} \partial^2_{ij} V +2 V_i \partial_t \partial_i V
\nonumber\\ &&\qquad~ +V \partial_t^2 V + \frac{3}{2} (\partial_t V)^2
- 2 \partial_i V_j \partial_j V_i\Bigr]\,,\label{X} \\ \hat{R}_i &=&
\Box^{-1}_\mathcal{R}\Bigl[- 4\pi G \bigl(\sigma_i V - \sigma
V_i\bigr) - 2 \partial_k V\partial_i V_k - \frac{3}{2} \partial_t V
\partial_i V \Bigr]\,,\label{Ri}\\ \hat{Z}_{ij} &=&
\Box^{-1}_\mathcal{R}\Bigl[- 4\pi G (\sigma_{ij}-\delta_{ij}
\sigma_{kk})V - 2\partial_{(i} V \partial_t V_{j)} \nonumber\\
&&\qquad~ + \partial_i V_k \partial_j V_k + \partial_k V_i \partial_k
V_j - 2 \partial_{(i} V_k \partial_k V_{j)} \nonumber\\ &&\qquad~ -
\delta_{ij} \partial_k V_m (\partial_k V_m - \partial_m V_k) -
\frac{3}{4} \delta_{ij} (\partial_t V)^2\Bigr]\,,\label{Zij}
\end{eqnarray}\end{subequations}}\noindent
together with the spatial traces denoted by $\hat{W}=\hat{W}_{ii}$ and ${\hat
Z}=\hat{Z}_{ii}$. Notice that we shall not need some higher-order potentials,
called $\hat{T}$ and $\hat{Y}_i$, which were crucial in the 3PN equations of
motion \cite{BFeom}. The 3PN multipole moments are in this sense ``less
non-linear'' than the 3PN equations of motion.

Like in paper I we find it convenient to decompose $I_L$ into three
pieces corresponding to the three terms in (\ref{IL}), respectively
referred to as ``scalar'' (S), ``vector'' (V) and ``tensor''
(T). Applying the formula (\ref{intdeltal}) we further decompose each
of these pieces into parts, labelled as I, II, III and so on,
according to the successive PN contributions. Hence, we write
\begin{eqnarray}\label{ILdecomp}
I_L&=&\mathrm{SI}_L+\mathrm{SII}_L+\mathrm{SIII}_L
+\mathrm{SIV}_L\nonumber\\&+&\mathrm{VI}_L
+\mathrm{VII}_L+\mathrm{VIII}_L\nonumber\\&+&\mathrm{TI}_L
+\mathrm{TII}_L\,,
\end{eqnarray}
in which we consistently neglect all terms that are higher-order than
3PN.\,\footnote{We generally do not indicate the PN remainder term
$\mathcal{O}\left(c^{-7}\right)$.} Without further comment and proof,
we give the explicit expressions of all these separate pieces, which
are equivalent to the similar expressions given by Eq. (4.2) in paper
I. Concerning the ``S-type'',
\allowdisplaybreaks{\begin{subequations}\label{Spart}\begin{eqnarray}
\mathrm{SI}_L &=& \mathop{\mathrm{FP}}_{B=0} \int d^3\mathbf{x}\,\vert
\widetilde{\mathbf{x}}\vert^B \hat{x}_L \biggl\{ \sigma -
\frac{1}{2\pi Gc^2} \Delta (V^2) + \frac{4V}{c^4} \sigma_{ii}
\nonumber \\ &&- \frac{2}{\pi Gc^4} V_i \partial_t \partial_i V -
\frac{1}{\pi Gc^4} \hat{W}_{ij}\partial^2_{ij}V - \frac{1}{2\pi Gc^4}
(\partial_t V)^2 + \frac{2}{\pi Gc^4} \partial_i V_j \partial_j V_i
\nonumber\\ &&- \frac{2}{3\pi Gc^4} \Delta (V^3) - \frac{1}{2\pi Gc^4}
\Delta (V {\hat W}) + \frac{16}{c^6} \sigma V_i V_i + \frac{8}{c^6}
\sigma_{ii} V^2 \nonumber\\ &&+ \frac{4}{c^6} \hat{W}_{ij} \sigma
_{ij} + \frac{1}{2\pi Gc^6} \hat{W} \partial^2_t V + \frac{1}{2\pi
Gc^6} V \partial^2_t \hat{W}\nonumber \\ &&- \frac{2}{\pi Gc^6} V
(\partial_t V)^2 - \frac{6}{\pi Gc^6} V_i \partial_t V \partial_i V -
\frac{4}{\pi Gc^6} V V_i \partial_t \partial_i V \nonumber\\ &&-
\frac{8}{\pi Gc^6} V_i \partial_j V_i \partial_j V + \frac{2}{\pi
Gc^6} (\partial_t V_i)^2 +\frac{1}{\pi Gc^6} \partial_t \hat{W}
\partial_t V \nonumber\\ &&+ \frac{4}{\pi Gc^6} \partial_i V_j
\partial_t \hat{W}_{ij} - \frac{4}{\pi Gc^6} \hat{Z}_{ij}
\partial_{ij} V - \frac{4}{\pi Gc^6}\partial_t \partial_i V \hat{R}_i
\nonumber\\ &&+ \frac{8}{\pi Gc^6} \partial_i V_j \partial_j \hat{R}_i
- \frac{2}{3\pi Gc^6} \Delta (V^4) - \frac{1}{\pi Gc^6} \Delta (V^2
\hat{W}) \nonumber\\ &&- \frac{1}{4\pi Gc^6} \Delta (\hat{W}^2 ) +
\frac{1}{2\pi Gc^6} \Delta (\hat{W}_{ij} \hat{W}_{ij}) - \frac{4}{\pi
Gc^6} \Delta (V \hat{X}) - \frac{2}{\pi Gc^6} \Delta (V \hat{Z})
\biggr\}\,,\label{SIpart}\\ \mathrm{SII}_L &=& \frac{1}{2c^2(2\ell
+3)} \mathop{\mathrm{FP}}_{B=0}\, \frac{d^2}{d t^2}\int
d^3\mathbf{x}\,\vert \widetilde{\mathbf{x}}\vert^B \biggl\{ \vert
\mathbf{x}\vert^2 \hat{x}_L \biggl[\sigma + \frac{4V}{c^4} \sigma_{ii}
- \frac{2}{\pi Gc^4} V_i \partial_t \partial_i V \nonumber\\ &&-
\frac{1}{\pi Gc^4} \hat{W}_{ij}\partial_{ij} V - \frac{1}{2\pi Gc^4}
(\partial_t V)^2 + \frac{2}{\pi Gc^4} \partial_i V_j \partial_j V_i
\biggr] \nonumber\\ &&- \frac{2\ell+3}{\pi Gc^2} \hat{x}_L V^2 -
\frac{1}{2\pi Gc^2} \partial_i \Bigl[\partial_i (V^2) \vert
\mathbf{x}\vert^2 \hat{x}_L - V^2 \partial_i (\vert \mathbf{x}\vert^2
\hat{x}_L)\Bigr] \nonumber \\ &&- \frac{2\ell+3}{\pi Gc^4} \hat{x}_L V
\hat{W} - \frac{1}{2\pi Gc^4} \partial_i \Bigl[\partial_i (V \hat{W})
\vert \mathbf{x}\vert^2 \hat{x}_L - V \hat{W} \partial_i (\vert
\mathbf{x}\vert^2 \hat{x}_L)\Bigr] \nonumber\\ &&-
\frac{4(2\ell+3)}{3\pi Gc^4} \hat{x}_L V^3 - \frac{2}{3\pi Gc^4}
\partial_i \Bigl[\partial_i (V^3) \vert \mathbf{x}\vert^2 \hat{x}_L -
V^3 \partial_i ( \vert \mathbf{x}\vert^2 \hat{x}_L)\Bigr]
\biggr\}\,,\label{SIIpart}\\ \mathrm{SIII}_L &=& \frac{1}{8c^4 (2\ell
+3)(2\ell +5)}\mathop{\mathrm{FP}}_{B=0}\, \frac{d^4}{d t^4}\int
d^3\mathbf{x}\,\vert \widetilde{\mathbf{x}}\vert^B \biggl\{ \vert
\mathbf{x}\vert^4 \hat{x}_L \sigma\nonumber\\ &&-
\frac{2(2\ell+5)}{\pi Gc^2}\vert \mathbf{x}\vert^2 \hat{x}_L V^2 -
\frac{1}{2\pi Gc^2}\partial_i \Bigl[\partial_i (V^2) \vert
\mathbf{x}\vert^4 \hat{x}_L - V^2 \partial_i (\vert \mathbf{x}\vert^4
\hat{x}_L)\Bigr] \biggr\}\,,\\ \mathrm{SIV}_L &=& \frac{1}{48c^6
(2\ell +3)(2\ell +5)(2\ell +7)} \mathop{\mathrm{FP}}_{B=0}\,
\frac{d^6}{d t^6}\int d^3\mathbf{x}\,\vert
\widetilde{\mathbf{x}}\vert^B \vert \mathbf{x}\vert^6
\hat{x}_L\sigma\,.
\end{eqnarray}\end{subequations}}\noindent
Then, the vectorial V-parts are
\allowdisplaybreaks{\begin{subequations}\label{Vpart}\begin{eqnarray}
\mathrm{VI}_L &=& -\frac{4 (2\ell +1)}{c^2 (\ell +1)(2\ell +3)}
\mathop{\mathrm{FP}}_{B=0}\, \frac{d}{d t}\int d^3\mathbf{x}\,\vert
\widetilde{\mathbf{x}}\vert^B \hat{x}_{iL} \biggl\{ \sigma_i +
\frac{2}{c^2} \sigma_i V - \frac{2}{c^2} \sigma V_i \nonumber \\ &&+
\frac{1}{\pi Gc^2} \partial_j V \partial_i V_j + \frac{3}{4\pi Gc^2}
\partial_t V \partial_i V - \frac{1}{2\pi Gc^2} \Delta (V V_i) +
\frac{2}{c^4} \sigma_i V^2 \nonumber \\ &&- \frac{4}{c^4} \sigma
\hat{R}_i + \frac{2}{c^4} V_i \sigma_{jj} + \frac{2}{c^4} \hat{W}_{ij}
\sigma_j + \frac{2}{c^4} V_j \sigma_{ij} + \frac{1}{2\pi Gc^4} V_i
\partial_t^2 V \nonumber \\ &&- \frac{1}{2\pi Gc^4} V \partial_t^2 V_i
+ \frac{1}{\pi Gc^4} \partial_t V \partial_t V_i - \frac{2}{\pi Gc^4}
V_j \partial_j \partial_t V_i \nonumber \\ &&+ \frac{3}{2\pi Gc^4} V
\partial_t V \partial_i V - \frac{1}{\pi Gc^4} V_i \partial_j V
\partial_j V + \frac{3}{2\pi Gc^4} V_j \partial_i V \partial_j V
\nonumber \\ &&+ \frac{2}{\pi Gc^4} \partial_j V \partial_i \hat{R}_j
- \frac{1}{\pi Gc^4} \hat{W}_{jk}\partial_{jk}^2 V_i + \frac{1}{\pi
Gc^4} \partial_t \hat{W}_{ij} \partial_j V \nonumber \\ &&-
\frac{1}{\pi Gc^4} \partial_j V_k \partial_i \hat{W}_{jk} +
\frac{1}{\pi Gc^4} \partial_j \hat{W}_{ik} \partial_k V_j -
\frac{1}{2\pi Gc^4} \Delta (V^2 V_i) \nonumber\\ &&- \frac{1}{\pi
Gc^4} \Delta (V \hat{R}_i) - \frac{1}{2\pi Gc^4} \Delta (\hat{W} V_i)
+ \frac{1}{2\pi Gc^4} \Delta (\hat{W}_{ij} V_j)\biggr\}\,,\\
\mathrm{VII}_L &=& -\frac{2(2\ell +1)}{c^4(\ell +1)(2\ell +3)(2\ell
+5)}\mathop{\mathrm{FP}}_{B=0}\, \frac{d^3}{d t^3}\int
d^3\mathbf{x}\,\vert \widetilde{\mathbf{x}}\vert^B \biggl\{ \vert
\mathbf{x}\vert^2 \hat{x}_{iL} \biggl[ \sigma_i \nonumber\\ &&+
\frac{2}{c^2} \sigma_i V - \frac{2}{c^2} \sigma V_i + \frac{1}{\pi
Gc^2} \partial_j V \partial_i V_j + \frac{3}{4\pi Gc^2} \partial_t V
\partial_i V \biggr] \nonumber \\ &&- \frac{2\ell+5}{\pi Gc^2}
\hat{x}_{iL} V V_i - \frac{1}{2\pi Gc^2} \partial_j \Bigl[\partial_j
(V V_i) \vert \mathbf{x}\vert^2 \hat{x}_{iL} - V V_i \partial_j (\vert
\mathbf{x}\vert^2 \hat{x}_{iL})\Bigr]\biggr\}\,,\\ \mathrm{VIII}_L &=&
-\frac{2\ell +1}{2c^6(\ell +1)(2\ell +3)(2\ell +5) (2\ell
+7)}\mathop{\mathrm{FP}}_{B=0}\, \frac{d^5}{d t^5}\int
d^3\mathbf{x}\,\vert \widetilde{\mathbf{x}}\vert^B \hat{x}_{iL} \vert
\mathbf{x}\vert^4 \sigma_i\,.
\end{eqnarray}\end{subequations}}\noindent
Finally the tensor parts read
\allowdisplaybreaks{\begin{subequations}\label{Tpart}\begin{eqnarray}
\mathrm{TI}_L &=& \frac{2(2\ell +1)}{c^4(\ell +1)(\ell +2)(2\ell
+5)}\mathop{\mathrm{FP}}_{B=0}\, \frac{d^2}{d t^2}\int
d^3\mathbf{x}\,\vert
\widetilde{\mathbf{x}}\vert^B\hat{x}_{ijL}\biggl\{ \sigma_{ij}
\nonumber\\ &&+ \frac{1}{4\pi G} \partial_i V \partial_j V +
\frac{4}{c^2} \sigma_{ij} V - \frac{4}{c^2} \sigma_i V_j +
\frac{2}{\pi Gc^2} \partial_i V \partial_t V_j \nonumber \\ &&-
\frac{1}{\pi Gc^2} \partial_i V_k \partial_j V_k + \frac{2}{\pi Gc^2}
\partial_i V_k \partial_k V_j - \frac{1}{2\pi Gc^2} \Delta (V_i V_j)
\biggr\}\,,\\ \mathrm{TII}_L &=& \frac{2\ell +1}{c^6 (\ell +1)(\ell
+2)(2\ell +5)(2\ell +7)}\mathop{\mathrm{FP}}_{B=0}\, \frac{d^4}{d
t^4}\int d^3\mathbf{x}\,\vert \widetilde{\mathbf{x}}\vert^B
\hat{x}_{ijL} \vert \mathbf{x}\vert^2 \biggl\{ \sigma_{ij}
\nonumber\\&&+ \frac{1}{4\pi G} \partial_i V \partial_j V \biggr\}\,.
\end{eqnarray}\end{subequations}}\noindent
These formulae \textit{stricto sensu} are valid for general
time-varying multipole moments having $\ell\geq 2$. However they
constitute also the main contributions in the conserved monopole and
dipole moments ($\ell=0,1$) as well. In fact we shall prove in Section
\ref{MD} that Eqs. (\ref{Spart})--(\ref{Tpart}) already give the
correct answer for the 3PN mass dipole moment of point particle
binaries, \textit{i.e.} $M_i=I_i$ and the quantity $\delta I_i$ given
in Eq. (\ref{DeltaIicons}) is zero at 3PN order.

\section{Hadamard regularization of the multipole moments}\label{Had}

We now specialize the general expression of the 3PN mass moments to
compact binary systems modelled by point particles. To this end the
first task is to compute all the necessary potentials
$\{V,\,V_i,\,\hat{W}_{ij},\,\cdots\}$, in the case of delta-function
singularities, using Hadamard's regularization. Actually the
computation of all these potentials has already been done at the
occasion of the 3PN equations of motion, and we refer to \cite{BFeom}
for the details. Our next task is to insert these potentials, and
their space-time derivatives, into
Eqs. (\ref{ILdecomp})--(\ref{Tpart}) for the quadrupole and dipole
moments, following the prescriptions of the Hadamard or more precisely
the pure-Hadamard-Schwartz (pHS) regularization.

\subsection{Pure Hadamard Schwartz regularization}\label{pHS}

Let us first recall the two concepts that constitute the basis of the
``ordinary'' Hadamard regularization
\cite{Hadamard,Schwartz}.\,\footnote{We refer to \cite{BDE04} for a
digest of possible variants of Hadamard's regularizatioon.} The first
one concerns the \textit{partie finie of a singular function} at the
value of a singular point. The generic function we have to deal with
reads $F(\mathbf{x})$, where $\mathbf{x}\in\mathbb{R}^3$, and becomes
singular at the two point-particle singularities located at the
positions $\mathbf{y}_1$ and $\mathbf{y}_2$ (in the harmonic
coordinate system). The function $F(\mathbf{x})$ is smooth
($C^\infty$) except at $\mathbf{y}_1$ and $\mathbf{y}_2$, and admits
around these singularities some Laurent-type expansions in powers of
$r_1\equiv \vert\mathbf{x}-\mathbf{y}_1\vert$ or $r_2\equiv
\vert\mathbf{x}-\mathbf{y}_1\vert$. When $r_1\rightarrow 0$ we have,
$\forall N\in\mathbb{N}$,
\begin{equation}\label{Fx}
F(\mathbf{x})=\sum_{p_0\leq p\leq N}r_1^p
\mathop{f}_1{}_p(\mathbf{n}_1)+o(r_1^N)\,,
\end{equation}
where the Landau $o$-symbol takes its usual meaning, and the
coefficients ${}_1f_p(\mathbf{n}_1)$ are functions of the unit vector
$\mathbf{n}_1\equiv (\mathbf{x}-\mathbf{y}_1)/r_1$.\,\footnote{For
clearer reading, we use a left-side label 1 like in
${}_1f_p(\mathbf{n}_1)$ when the quantity appears within the text,
however the label is always put underneath the quantity when it
appears in an equation like in (\ref{Fx}).} We have $p\in \mathbb{Z}$,
bounded from below by some typically negative integer $p_0$ depending
on the $F$ in question. The class of functions such as $F$ is called
$\mathcal{F}$; see Ref. \cite{BFreg} for a fuller account of the
properties of functions in this class. Now the Hadamard partie finie
of $F$ at the singular point $\mathbf{y}_1$, denoted $(F)_1$, is
defined by the angular average
\begin{equation}\label{F1}
(F)_1 \equiv\int
\frac{d\Omega_1}{4\pi}\mathop{f}_1{}_0(\mathbf{n}_1)\,,
\end{equation}
where $d\Omega_1\equiv d\Omega(\mathbf{n}_1)$ is the solid angle
element on the unit sphere centered on $\mathbf{y}_1$. Note that the
spherical average (\ref{F1}) is performed in a \textit{global}
inertial frame. In the context of the extended-Hadamard regularization
\cite{BFregM} one defines the regularization (\ref{F1}) in the
Minkowskian ``rest frame'' of each particle. A distinctive feature of
the partie finie (\ref{F1}) is its ``non-distributivity'' in the sense
that
\begin{equation}\label{nondistr}
(FG)_1\not=(F)_1(G)_1~~\text{in general for $F,G\in\mathcal{F}$.}
\end{equation}

The second notion in Hadamard's regularization is that of the
\textit{partie finie of a divergent integral}, which attributes a
value to the integral over $\mathbb{R}^3$ of the function
$F(\mathbf{x})$. Consider first two ``regularization volumes'' around
the two singularities $\mathbf{y}_1$ and $\mathbf{y}_2$. We can
specifically choose two \textit{spherical} balls (in the considered
coordinate system), $B_1(s)$ and $B_2(s)$, centered on the
singularities, each of them with radius $s$. The integral of $F$ over
the domain exterior to these balls, \textit{i.e.}
$\mathbb{R}^3\setminus B_1(s)\cup B_2(s)$, is well-defined for any
$s>0$. Hadamard's partie finie (Pf) of the generally divergent
integral of $F$ is then \textit{defined} by the always existing limit
\begin{eqnarray}\label{Pf}
\text{Pf}_{s_1,s_2}\int
d^3\mathbf{x}\,F(\mathbf{x})&\equiv&\lim_{s\rightarrow
0}\,\biggl\{\int_{\mathbb{R}^3\setminus B_1(s)\cup
B_2(s)}d^3\mathbf{x}\,F(\mathbf{x})\nonumber\\ &&+ \,4\pi
\sum_{p+3<0}\frac{s^{p+3}}{p+3}\biggl(\frac{F}{r_1^p}\biggr)_1 +
4\pi\,\ln \left(\frac{s}{s_1}\right)\bigl(r_1^3
F\bigr)_1+1\leftrightarrow 2\biggr\}.
\end{eqnarray}
The extra terms, which involve some parties finies in the sense of
(\ref{F1}), are such that they cancel out the singular part of the
``exterior'' integral when $s\rightarrow 0$. Here the symbol
$1\leftrightarrow 2$ means the same terms but corresponding to the
other particle. The two constants $s_1$ and $s_2$ entering the
logarithmic terms of this definition play a very important role at 3PN
order. A way to interpret them is to say that they reflect the
arbitrariness in the choice of the regularization volumes surrounding
the particles. Indeed it can be checked that the Hadamard partie finie
(\ref{Pf}) does not depend, modulo changing the values of $s_1$ and
$s_2$, on the \textit{shape} of $B_1$ and $B_2$, above chosen as
simple spherical balls (see the discussion in \cite{BFreg}).

The two notions of partie finie, (\ref{F1}) and (\ref{Pf}), are
intimately related. Notably the partie-finie integral (\ref{Pf}) of a
gradient is in general non-zero but given by the partie finie, in the
sense of (\ref{F1}), of some singular function (see \cite{BFreg} for
more details). With the definitions (\ref{F1}) and (\ref{Pf}) one can
show that, if we want to dispose of a \textit{local} meaning (at any
field point $\mathbf{x}$) for the product of $F$ with a
delta-function, say $F(\mathbf{x})\,\delta(\mathbf{x}-\mathbf{y}_1)$,
then one cannot simply replace $F$ in front of the delta-function by
its regularized value. This is a consequence of the non-distributivity
of Hadamard's partie finie, Eq. (\ref{nondistr}). Thus,
\begin{equation}
F(\mathbf{x})\,\delta(\mathbf{x}-\mathbf{y}_1) \not=(F)_1\,\delta(
\mathbf{x}-\mathbf{y}_1)~~\text{in general for $F\in\mathcal{F}$.}
\label{deltanondistr}
\end{equation}

It is quite evident that the two properties (\ref{nondistr}) and
(\ref{deltanondistr}) are problematic. A remarkable fact is that the
problem of the non-distributivity, Eqs. (\ref{nondistr}) and
(\ref{deltanondistr}), arises precisely at the 3PN order, for both the
radiation field and the equations of motion, and not before that
order. In the problem of the equations of motion we could deal with
the properties (\ref{nondistr}) and (\ref{deltanondistr}) by
implementing the extended Hadamard regularization of
Refs. \cite{BFreg,BFregM}. We have not (yet) succeeded in applying the
extended Hadamard regularization to the problem of gravitational wave
generation. For the present paper we choose to follow a different
route, and adopt the \textit{pure-Hadamard-Schwartz} (pHS)
regularization defined in Ref. \cite{BDE04}.

The pHS regularization is a specific ``minimal'' variant of the
Hadamard regularization, which is designed in such a way that it
avoids, \textit{by its very definition}, the problematic consequences
(\ref{nondistr}) and (\ref{deltanondistr}) of the ``ordinary''
Hadamard regularization. It applies to the case relevant here where
the singular function, say $F_L\in\mathcal{F}$, is made of sums of
products of the non-linear potentials $V$, $V_i$, $\hat{W}_{ij}$,
$\cdots$ and their space-time derivatives $\partial_i V$, $\cdots$,
and is multiplied by some (regular) multipolar factor $\hat{x}_L$,
that is
\begin{equation}\label{FcalP}
F_L(\mathbf{x})=\hat{x}_L\,\mathcal{P} [ V, \,V_i, \,\hat{W}_{ij},
\,\cdots, \,\partial_i V, \cdots]\,,
\end{equation}
where $\mathcal{P}$ denotes a certain multilinear form, \textit{i.e.}
a polynomial in each of its variables $V, \,V_i, \,\cdots$. The rules
of the pHS regularization are: (i) an integral $\int d^3 \mathbf{x} \,
F_L(\mathbf{x})$, where $F_L$ takes the form (\ref{FcalP}), is
regularized according to the partie finie prescription (\ref{Pf}) like
in the ordinary Hadamard regularization; (ii) we add the contribution
of \textit{distributional} terms coming from the derivatives of
potentials, $\partial_i V, \cdots$, according to the usual Schwartz
distribution theory \cite{Schwartz} (this point is detailed in Section
\ref{schwartz}); (iii) the regularization of a product of potentials
$V$, $V_i$, $\hat{W}_{ij}$, $\cdots$ (and their gradients) at a
singular point is assumed to be ``distributive'', which means that the
value of $F_L$ at point 1 (say) is given by the replacement rule
\begin{equation}\label{calF1}
(F_L)_1~\longrightarrow~ \hat{y}_1^L\,\mathcal{P} [ (V)_1, \,(V_i)_1,
\,(\hat{W}_{ij})_1, \,\cdots, \,(\partial_i V)_1, \,\cdots]\,,
\end{equation}
where the partie finie (\ref{F1}) is applied individually on each of
the potentials, and $\hat{y}_1^L=\mathrm{STF}(y_1^{i_1}\cdots
y_1^{i_\ell})$; and (iv) a ``contact'' term, \textit{i.e.} of the form
$F_L(\mathbf{x})\,\delta(\mathbf{x}-\mathbf{y}_1)$, appearing in the
calculation of the sources of the non-linear potentials and
corresponding to their ``compact-support'' parts, is regularized by
means of the rule
\begin{equation}\label{calFdelta}
F_L(\mathbf{x})\,\delta(\mathbf{x}-\mathbf{y}_1)~\longrightarrow~
\hat{y}_1^L\,\mathcal{P} [ (V)_1, \,(V_i)_1, \,(\hat{W}_{ij})_1,
\,\cdots]\,\delta( \mathbf{x}-\mathbf{y}_1)\,.
\end{equation}
The rules (\ref{calF1}) and (\ref{calFdelta}) of the pHS
regularization are well defined, and are not submitted, by definition,
to the unwanted consequences of the non-distributivity of the
\textit{ordinary} Hadamard regularization: (\ref{nondistr}) and
(\ref{deltanondistr}). However, as we shall emphasize in Section
\ref{pointpart}, the pHS regularization becomes \textit{physically
incomplete} at the 3PN order, in the sense that it must be augmented
by certain ambiguous contributions, which \textit{a priori} cannot be
determined within this regularization scheme.

Our motivation for introducing the pHS regularization is that it
constitutes in some sense the \textit{core} of both the Hadamard and
dimensional regularizations \cite{BDE04,BDEI04}. By ``core'' we mean
that it will yield the complete and correct result for all the terms
\textit{but for a few}, and for those which cannot be determined
unambiguously the undetermined part will take in general a very
special and limited type of structure. Hence the correct result is
obtained by adding to the pHS result a limited number of ``ambiguous''
terms, parametrized by some arbitrary numerical coefficients called
ambiguity parameters.

In dimensional regularization the undetermined terms correspond
exactly to the contribution of \textit{poles} $\propto 1/\varepsilon$,
where $d=3+\varepsilon$ is the dimension of space. The complete result
in dimensional regularization appears therefore as the sum of the pHS
result and what we call the ``difference'', namely the pole part $\sim
1/\varepsilon$ which can be quite easily obtained from the expansion
near the singularities of the functions involved \cite{BDE04,BDEI04},
and which is nothing but the difference between the dimensional and
the pHS regularizations. The method for determining the ambiguity
parameters is therefore to equate the ambiguous terms, as they are
defined with respect to the pHS regularization, to the latter
difference. In the present paper, we compute the pHS regularization of
the 3PN binary's quadrupole moment; this constitutes the first and
necessary step toward the complete calculation by dimensional
regularization (see Ref. \cite{BDEI04} for a summary of the method).

\subsection{Schwartz distributional derivatives}\label{schwartz}

We detail here an important feature of the pHS regularization, namely
the systematic use of distributional derivatives \textit{\`a la}
Schwartz \cite{Schwartz}. Recall first that previous work on the
equations of motion \cite{BFeom} showed that the Schwartz
distributional derivatives yield \textit{ill-defined} (formally
infinite) terms at the 3PN order in ordinary three-dimensional
space. This was a motivation for introducing some appropriate
generalized versions of distributional derivatives in the context of
the extended Hadamard regularization \cite{BFreg}. However, one can
show \cite{BDE04} that, by working in a space with $d$ dimensions
instead of 3 dimensions, and invoking complex analytic continuation in
$d$, the latter ill-defined terms are in fact rigorously
\textit{zero}. The usual Schwartz distributional derivatives are
therefore well-defined in the context of \textit{dimensional
regularization}, and they have been shown to contribute in an
essential way to the final equations of motion at 3PN order
\cite{BDE04}.

In the present paper we include the Schwartz distributional
derivatives as part of the calculation based on the pHS
regularization. However, as we just pointed out the Schwartz
derivatives yield ill-defined terms in 3 dimensions, so we shall
compute them in $d$ spatial dimensions, and then \textit{take the
limit}
\begin{equation}\label{eps0}
\varepsilon\equiv d-3~\rightarrow~0\,.
\end{equation}
This permits us to cancel out (by dimensional continuation) all the
formally divergent terms and to perform a perfectly rigorous
calculation. Of course, this way of handling the Schwarztian
distributional terms shows that in fact the calculation will already
constitute a part of a complete calculation using dimensional
regularization. However the spirit is different. In the present
calculation we use dimensional continuation as a mathematical trick
enabling us to give a well-defined meaning to a limited number of
terms which would be otherwise infinite. In a real computation based
on dimensional regularization the scope is broader, and we should
start from the Einstein field equations in $d$ dimensions and
consistently perform all the derivations for arbitrary
$d\in\mathbb{C}$ before eventually taking the limit (\ref{eps0}). In
the present paper we shall perform our calculation of the Schwartz
derivatives based on the expression for the multipole moment given by
Eqs. (\ref{Spart})-(\ref{Tpart}), \textit{i.e.}  without taking into
account the modification of the various coefficients which would come
from the Einstein field equations in $d$ dimensions. The result in the
limit $\varepsilon\rightarrow 0$ will however exactly be the same as
by including such $d$-dependent coefficients because the
distributional parts of Schwartz derivatives do not generate any poles
proportional to $1/\varepsilon$.

One may ask why is it possible to choose, in the context of Hadamard's
regularization, different prescriptions for the distributional
derivatives, and nevertheless obtain the same physical result at the
end\,? For instance what would happen if instead of using the Schwartz
distributional derivative in the way we have just described, we adopt
the generalized derivatives of the extended Hadamard regularization
\cite{BFreg}\,? The answer which emerges from our detailed
computations is that the difference between the final results obtained
by different prescriptions takes the form of the ``ambiguous'' terms,
which are given in the case of the quadrupole moment by the R.H.S. of
Eq. (\ref{Iijres}) below. Thus, different calculations are in fact
equivalent modulo some simple redefinition (or shift) of the values of
the ambiguity parameters $\xi$, $\kappa$ and $\zeta$.

The $d$-dimensional calculation of the Schwartz distributional
derivatives in the 3PN moments essentially necessitates the same
ingredients as in the problem of equations of motion \cite{BDE04}. We
introduce some elementary Poisson kernels $u_1$ and $v_1$, solving the
equations
\begin{subequations}\label{Deltau1v1}\begin{eqnarray}
\Delta u_1 &=& -4 \pi\,\delta^{(d)}(\mathbf{x}-\mathbf{y}_1)\,,\\
\Delta v_1 &=& u_1\,,
\end{eqnarray}\end{subequations} 
where $\Delta$ is Laplace's operator in $d$ dimensions and
$\delta^{(d)}$ is the Dirac delta-function in $d$ dimensions. These
kernels play a crucial role in the construction of the $d$-dimensional
versions of the non-linear potentials \cite{BDE04}. They parametrize
the compact-support potential $V$ at 1PN order; evidently $u_1$ enters
the Newtonian part of $V$ while the twice-iterated Poisson kernel
$v_1$ is used for the 1PN retardation. The kernels are given by
\begin{subequations}\label{u1v1}\begin{eqnarray}
u_1&=&\tilde{k}\,r_1^{2-d}\,,\\
v_1&=&\frac{\tilde{k}\,r_1^{4-d}}{2(4-d)}\,,
\end{eqnarray}\end{subequations}
where $r_1\equiv\vert\mathbf{x}-\mathbf{y}_1\vert$ and $\tilde{k}$ is
related to the Eulerian $\Gamma$-function by
\begin{subequations}\label{ktilde}\begin{eqnarray}
\tilde{k}&=&\frac{\Gamma\left(\frac{d-2}{2}\right)}{\pi^{
\frac{d-2}{2}}}\,,\\ \lim_{d\rightarrow 3}\tilde{k}&=&1\,.
\end{eqnarray}\end{subequations}
The \textit{second} partial derivative of $u_1$, and the
\textit{fourth} partial derivative of $v_1$, will contain, besides an
ordinary singular function (or pseudo-function) obtained by performing
the derivative in an ``ordinary'' sense, a distributional component
proportional to $\delta^{(d)}$, and given by
\begin{subequations}\label{du1v1}\begin{eqnarray}
\partial_{ij}^2 (u_1)&=&\partial_{ij}^2
(u_1)_{\bigl|_\text{ordinary}}-\frac{4
\pi}{d}\,\delta_{ij}\,\delta^{(d)} (\mathbf{x}-\mathbf{y}_1)\,,\\
\partial_{ijkl}^4 (v_1)&=&\partial_{ijkl}^4
(v_1)_{\bigl|_\text{ordinary}}
-\frac{4\pi}{d(d+2)}\Bigl(\delta_{ij}\delta_{kl}+\delta_{ik}
\delta_{jl}+\delta_{il}\delta_{jk}\Bigr)\,\delta^{(d)} (\mathbf{x}
-\mathbf{y}_1)\,.
\end{eqnarray}\end{subequations} 
These expressions can be derived as particular cases of the
Gel'fand-Shilov formula \cite{gelfand}. In addition, we can treat the
distributional \textit{time}-derivatives in a very simple way from the
rule $\partial_t=-v_1^i\,\partial_i$ applicable to the purely
distributional part of the derivative. 

The expressions (\ref{du1v1}) permit for instance the computation of
the distributional derivative $\partial_{ij}^2V$ at the 1PN level to
be inserted into the 1PN source term
$\sim\hat{W}_{ij}\partial_{ij}^2V$ in the expression of $I_L$,
\textit{cf.} Eq. (\ref{SIpart}). Let us emphasize that the previous
method of introducing the Schwartz distributional derivatives in the
pHS formalism, \textit{i.e.} by means of dimensional continuation in
$d$, is probably the only rigorous way to do it. An alternative
approach would consist of staying in 3 dimensions, and employing the
generalized derivative operators defined in \cite{BFreg} (they act on
singular functions of the class $\mathcal{F}$ instead of smooth
``test'' functions with compact support as in Schwartz's
distributional theory). But then the result will differ from
Schwartz's derivatives by some terms having the structure of the
ambiguous terms in the R.H.S. of Eq. (\ref{Iijres}).

\subsection{Three-dimensional partie finie integrals}\label{HadPf}

The main basis of the computation of the 3PN multipole moments of
point particles is to perform explicitly many three-dimensional
non-compact support integrals in the sense of the Hadamard partie
finie (\ref{Pf}). In addition to the partie finie we have to take care
of the finite part process based on analytic continuation in
$B\in\mathbb{C}$ to deal with the boundary of the integrals at
infinity. Therefore we must compute explicitly many integrals of the
type
\begin{equation}\label{I}
\mathcal{I}[s_1,s_2,r_0] \equiv
\mathop{\mathrm{FP}}_{B=0}\bigg\{\mathrm{Pf}_{s_1,s_2}\int d^3
\mathbf{x}~\vert\widetilde{\mathbf{x}}\vert^B F(\mathbf{x})\bigg\}\,,
\end{equation}
which depend \textit{a priori} on the two UV-type length scales $s_1$
and $s_2$ associated with the Hadamard partie finie (\ref{Pf}), and on
the IR-type length scale $r_0$ introduced into the general formalism
of Section \ref{multdecomp} through the regularization factor
$\vert\widetilde{\mathbf{x}}\vert^B\equiv\vert\mathbf{x}/r_0\vert^B$.

The function $F(\mathbf{x})$ in Eq. (\ref{I}) stands for a non-compact
support function which, as far as its UV properties are concerned,
belongs to the class of singular functions $\mathcal{F}$,
\textit{i.e.} admits some expansions of the type (\ref{Fx}). The IR
behavior of $F(\mathbf{x})$, when $\vert\mathbf{x}\vert\rightarrow
+\infty$, will be specified below. $F(\mathbf{x})$ contains also some
multipolar factor such as $\hat{x}_L$ but for simplicity's sake we do
not indicate here the multi-index $L$. In the general case
$F(\mathbf{x})$ admits an expression such as Eq. (\ref{FcalP}),
\textit{i.e.} it is given by some multi-linear functional of the
elementary potentials $V,\,V_i,\hat{W}_{ij},\,\cdots$ and their
derivatives. The function $F$ represents the sum of all the
\textit{non-compact support} terms in
Eqs. (\ref{Spart})--(\ref{Tpart}), taking into account only the
\textit{ordinary} parts of the derivatives. The compact support terms
in (\ref{Spart})--(\ref{Tpart}), as well as the purely distributional
parts of the Schwartz derivatives [calculated with (\ref{du1v1})], are
treated separately using the rules of the pHS regularization for
contact terms, see Eqs. (\ref{calF1})--(\ref{calFdelta}). On the other
hand, we shall point out in Section \ref{infbound} that for many
non-compact terms in Eqs. (\ref{Spart})--(\ref{Tpart}) we can after
integrating by parts perform a much simpler computation of these
terms, confined to the boundary of the integral ``at infinity'' and
depending on the sole properties of the finite part operation
$\mathop{\mathrm{FP}}_{B=0}$.

In this Section we explain our practical method for dealing with the
integral (\ref{I}). The basic idea is to relate (\ref{I}) to an
integral which is convergent at infinity, on which we can thus remove
the finite part at $B=0$, and then to compute this integral by means
of the very efficient method of ``angular integration'' described by
Eq. (4.17) in Ref. \cite{BFreg}. We assume for this calculation (this
will always be verified in practice) that $F$ admits a power-like
expansion \textit{at infinity}, when $r_1\rightarrow +\infty$ with
$t=\mathrm{const}$, of the following type (for any large enough $M$)
\begin{equation}\label{Finf}
F(\mathbf{x})=\sum_{k_0\leq k\leq M} \,\frac{1}{r_1^k}
\,\mathop{\varphi}_1{}_k(\mathbf{n}_1)+o\left(\frac{1}{r_1^M}\right)\,,
\end{equation} 
where the coefficients ${}_1\varphi_k$ depend on the unit vector
$\mathbf{n}_1=(\mathbf{x}-\mathbf{y}_1)/r_1$. The index $k$ is bounded
from below by some $k_0\in\mathbb{Z}$. For convenience we have singled
out the singularity 1, and considered the expansion when
$r_1\rightarrow +\infty$, instead of the more natural choice
$\vert\mathrm{x}\vert\rightarrow +\infty$. Introducing such an
asymmetry between the points 1 and 2 is only a matter of convenience,
but in fact it is quite appropriate in the present formalism because
we shall later use the method of ``angular integration'' \cite{BFreg}
which already particularizes the point 1, around which the angular
integration is performed. An advantage is that a good check of the
calculation can be done at the end since the final result will have to
be symmetric in the particle exchange $1\leftrightarrow 2$. Next we
define an auxiliary function ${}_1F_{\infty}$ by \textit{subtracting}
from $F$ all the terms in its expansion (\ref{Finf}) which yield some
divergencies at infinity, \textit{i.e.}
\begin{equation}\label{F1inf}
\mathop{F}_1{}_{\!\!\infty}(\mathbf{x}) \equiv F(\mathbf{x}) -
\sum_{k_0\leq k\leq 3} \,\frac{1}{r_1^k}
\,\mathop{\varphi}_1{}_k(\mathbf{n}_1)\,.
\end{equation}
The integral of ${}_1F_{\infty}$ is easily seen to be convergent at
infinity, and therefore it can be computed with the ordinary Hadamard
partie finie prescription given by (\ref{Pf}). Inserting (\ref{F1inf})
into (\ref{I}) we obtain
\begin{equation}\label{I1}
\mathcal{I}[s_1,s_2,r_0] = \text{Pf}_{s_1,s_2}\int
d^3\mathbf{x}\,\mathop{F}_1{}_{\!\!\infty} + \sum_{k_0\leq k\leq 3}
\mathop{\mathrm{FP}}_{B=0}\biggl\{\mathrm{Pf}_{s_1,s_2}\int d^3
\mathbf{x}~\vert\widetilde{\mathbf{x}}\vert^B\frac{1}{r_1^k}
\mathop{\varphi}_1{}_k(\mathbf{n}_1)\biggr\}\,.
\end{equation}

We transform the second term in the R.H.S. of (\ref{I1}). The
integrand in the curly brackets is replaced by the following
equivalent when $B\rightarrow 0$ limited at \textit{first order} in
$B$,
\begin{equation}\label{Iexp}
\int d^3 \mathbf{x}~\vert\widetilde{\mathbf{x}}\vert^B\frac{1}{r_1^k}
\mathop{\varphi}_1{}_k(\mathbf{n}_1) = \int d^3
\mathbf{r}_1~\widetilde{r}_1^B\left[1+B \ln\left(\frac{\vert\mathbf{x}
\vert}{r_1}\right)+\mathcal{O}\left(B^2\right)\right]\frac{1}{r_1^k}
\mathop{\varphi}_1{}_k(\mathbf{n}_1)\,.
\end{equation}
(Notice our change of integration variable, from $\mathbf{x}$ to
$\mathbf{r}_1=\mathbf{x}-\mathbf{y}_1$; we pose $\widetilde{r}_1\equiv
r_1/r_0$.) This will turn out to be sufficient for our purpose because
at 3PN order one can show that there are no multiple poles in $B$,
therefore the neglected terms $\mathcal{O}\left(B^2\right)$ will never
contribute at this order. We can prove that the first term in the
R.H.S. of (\ref{Iexp}) is zero except when $k=3$, in which case the
integral admits a pole, and its finite part depends on the
\textit{logarithm} of the ratio between $r_0$ and $s_1$. Furthermore,
the next term, carrying an explicit factor $B$, is found to be zero in
the case $k=3$, so we get
\begin{equation}\label{k3}
\mathop{\mathrm{FP}}_{B=0}\biggl\{\mathrm{Pf}_{s_1,s_2}\int d^3
\mathbf{x}~\vert\widetilde{\mathbf{x}}\vert^B\frac{1}{r_1^3}
\mathop{\varphi}_1{}_3(\mathbf{n}_1)\biggr\} =
\ln\left(\frac{r_0}{s_1}\right)\int
d\Omega_1\mathop{\varphi}_1{}_3(\mathbf{n}_1)\,.
\end{equation}
Consider next the generic cases where $k\leq 2$. It is clear that the
integrals are then convergent when $r_1\rightarrow 0$ so we can ignore
the Hadamard partie finie ($\mathrm{Pf}$). By analytic continuation in
$B$ we find that the first term in (\ref{Iexp}) is now zero, and it
remains the next one, which can clearly contribute only if the
integral develops a simple pole $\sim 1/B$ at infinity. When
$r_1\rightarrow +\infty$ we have the expansion
\begin{equation}\label{logexp}
\ln\left(\frac{\vert\mathbf{x}
\vert}{r_1}\right)=\frac{1}{2}\sum_{m=1}^{+\infty}\frac{
\alpha_m(\mathbf{n}_1)}{r_1^m}\,,
\end{equation}
where the various coefficients $\alpha_m$ depend on $\mathbf{n}_1$ and
also on $\mathbf{y}_1$, and are related to the Gegenbauer polynomial
$C_m^\mu (t)$ by\,\footnote{Here we follow the standard convention for
the Gegenbauer polynomial \cite{GZ}. In Eq. (\ref{alpham}) it is
calculated at the value $t =
-\frac{(\mathbf{n}_1\mathbf{y}_1)}{\vert\mathbf{y}_1\vert}$, where
$(\mathbf{n}_1\mathbf{y}_1)$ denotes the usual scalar product and
$\vert\mathbf{y}_1\vert$ the usual norm.}
\begin{equation}\label{alpham}
\alpha_m(\mathbf{n}_1) = -\vert\mathbf{y}_1\vert^m\left\{
\frac{d}{d\mu}\left[C_m^\mu\left(
-\frac{(\mathbf{n}_1\mathbf{y}_1)}{\vert\mathbf{y}_1
\vert}\right)\right]\right\}_{\mu = 0}\,,
\end{equation}
where one sets $\mu=0$ after differentiation of $C_m^\mu (t)$ with
respect to its argument $\mu$. One may want to express (\ref{alpham})
in a more detailed way with the help of Rodrigues' formula for the
Gegenbauer polynomial. Substituting the expansion
(\ref{logexp})--(\ref{alpham}) into Eq. (\ref{Iexp}) we are finally in
a position to obtain the looked-for result
\begin{eqnarray}\label{Ires}
\mathcal{I}[s_1,s_2,r_0] &=& \text{Pf}_{s_1,s_2}\int
d^3\mathbf{x}\,\mathop{F}_1{}_{\!\!\infty} \nonumber\\ &+&
\ln\left(\frac{r_0}{s_1}\right)\int
d\Omega_1\,\mathop{\varphi}_1{}_3(\mathbf{n}_1) -
\frac{1}{2}\sum_{m=1}^{+\infty}\int d\Omega_1 \,\alpha_m(\mathbf{n}_1)
\,\mathop{\varphi}_1{}_{3-m}(\mathbf{n}_1)\,.
\end{eqnarray}
This formula is systematically employed in our algebraic computer
programs. [Of course, the sum in the last term is in fact finite
because $3-m\geq k_0$; see Eq. (\ref{Finf}).] As we said the first
term (partie finie integral) is computed by means of an ``angular
integration'' around the particle 1 following the procedure defined by
(4.17) in Ref. \cite{BFreg}. We have found that the 3PN quadrupole
moment resulting from the systematic application of Eq. (\ref{Ires})
is in perfect agreement with the result of paper I, which was derived
by ``case-by-case'' integration, \textit{i.e.} using different methods
depending on the type and structure of the various terms encountered
in the problem.

\subsection{Contributions depending on the boundary at infinity}\label{infbound}

The result (\ref{Ires}) can be applied to any of the non-compact
support terms in (\ref{Spart})--(\ref{Tpart}). However, we now show
that many terms can be re-expressed, after suitable integration by
parts, in the form of a \textit{surface integral} at infinity
$r\equiv\vert\mathbf{x}\vert\rightarrow +\infty$. Evaluating the
surface integral is in general much simpler than performing the
``bulk'' calculation following Eq. (\ref{Ires}). The first type of
term in (\ref{Spart})--(\ref{Tpart}) for which a computation ``at
infinity'' is possible takes the form of the finite part
($\mathop{\mathrm{FP}}_{B=0}$) of an integral involving the product of
a multipolar STF factor $\hat{x}_L$ with the \textit{Laplacian} of
some $G\in\mathcal{F}$, having the structure of a product of
elementary potentials, \textit{i.e.}
\begin{equation}\label{J}
\mathcal{J}_L \equiv \mathop{ \mathrm{FP}}_{B=0}\,\int
d^3\mathbf{x}\,\vert \widetilde{\mathbf{x}}\vert^B\,\hat{x}_L\,\Delta
G\,.
\end{equation}
There are many such terms in (\ref{Spart})--(\ref{Tpart}), see for
instance the last six terms in Eq. (\ref{SIpart}). The second type of
term which is amenable to a treatment at infinity is composed of the
divergence of some vectorial function $H_i\in\mathcal{F}$, containing
itself some multipolar factor $\hat{x}_L$ (not indicated in our
notation for $H_i$), say
\begin{equation}\label{K}
\mathcal{K} \equiv \mathop{\mathrm{FP}}_{B=0}\,\int
d^3\mathbf{x}\,\vert \widetilde{\mathbf{x}}\vert^B\,\partial_iH_i\,.
\end{equation}
An example is given by the last two terms in Eq. (\ref{SIIpart}). We
deal with these two categories of terms, $\mathcal{J}_L$ and
$\mathcal{K}$, in turn.

The Laplacian in $\mathcal{J}_L$ is integrated by parts, and we are
allowed to cancel out the all-integrated term which is zero by
analytic continuation in $B\in\mathbb{C}$ (because it is zero in the
case where $\Re (B)$ is chosen to be a large enough \textit{negative}
number), thereby obtaining
\begin{equation}\label{Jint}
\mathcal{J}_L = \mathop{ \mathrm{FP}}_{B=0}\,\int
d^3\mathbf{x}\,\Delta\left(\vert
\widetilde{\mathbf{x}}\vert^B\,\hat{x}_L\right)G = \mathop{
\mathrm{FP}}_{B=0}\,B(B+2\ell+1)\int d^3\mathbf{x}\,\vert
\widetilde{\mathbf{x}}\vert^{B-2}\,\frac{\hat{x}_L}{r_0^2}\,G\,.
\end{equation}
The presence of the factor $B$ means that the result depends only on
the polar part $\sim 1/B$ of the integral at the boundary at
infinity. Since the pole comes exclusively from a radial integral of
the type $\int dr \,r^{B-1}=r^B/B$, we need only to look for the term
of the order of $r^{-\ell-1}$ in the expansion of $G$ when
$r\rightarrow +\infty$. We compute the expansion of $G$ and obtain
\begin{equation}\label{Gexp}
G = \cdots + \frac{1}{r^{\ell+1}} \,X_\ell (\mathbf{n}) +
\mathcal{O}\left(\frac{1}{r^{\ell+2}}\right)\,,
\end{equation}
where the dots indicate the terms having different magnitudes in $1/r$
and which thus do not concern us for the present calculation. The
interesting coefficient in (\ref{Gexp}) is $X_\ell (\mathbf{n})$, and
in terms of it we get
\begin{equation}\label{Jpole} 
\mathcal{J}_L = \mathop{
\mathrm{FP}}_{B=0}\,B(B+2\ell+1)\,r_0^{-B}\int_\mathcal{R}^{+\infty} d
r \,r^{B-1} \int d \Omega \, \hat{n}_L \,X_\ell (\mathbf{n})\,,
\end{equation}
in which we indicated that the radial integral depends only on a
neighborhood of infinity, from some arbitrary radius $\mathcal{R}$ up
to $+\infty$. This point is actually not completely obvious at this
stage and must be justified in the following way. In the general
formalism of Ref. \cite{B98mult}, which is valid for extended
\textit{smooth} matter distributions, any integral having a factor $B$
in front will depend only the behavior of the integrand at
infinity. Indeed since the matter source is smooth the near zone part
of the integral is convergent, thus no poles $\propto 1/B$ can arise
due to the UV behavior of the integrand and only the IR-type poles can
contribute to the value of the integral. When applying the formalism
to point particles one must keep this feature in mind, and replace the
stress-energy tensor of an extended source by the $T^{\mu\nu}$ of
point-particles in the term in question already in the form, by the
previous argument, of some far-zone integral. Thus, even for point
particles the term depends only on the boundary at infinity and does
not explicitly involve UV-type divergencies, although it may
implicitly contain some contributions coming from UV divergencies
occuring at previous PN iteration steps. Finally, from
Eq. (\ref{Jpole}) we readily find the result
\begin{equation}\label{Jfinal} 
\mathcal{J}_L = -(2\ell+1)\int d \Omega \, \hat{n}_L \,X_\ell
(\mathbf{n})\,.
\end{equation}
We notice that this result is independent of the arbitrary scale
$\mathcal{R}$ introduced in (\ref{Jpole}), as well as of the IR
constant $r_0$. Concerning the integral $\mathcal{K}$ defined by
Eq. (\ref{K}) we proceed similarly by integration by parts. We find
that the term depends only on the part in the expansion of $H_i$ at
infinity which goes like $1/r^2$, hence we look for the coefficient
$Y_i (\mathbf{n})$ in
\begin{equation}\label{Hiexp}
H_i = \cdots + \frac{1}{r^2} \,Y_i (\mathbf{n}) +
\mathcal{O}\left(\frac{1}{r^3}\right)\,,
\end{equation}
and we obtain the simple result (independent of $\mathcal{R}$ and
$r_0$)
\begin{equation}\label{Kfinal}
\mathcal{K} = \int d \Omega \, n_i \,Y_i (\mathbf{n})\,.
\end{equation}

In summary, many non-compact support terms in
(\ref{Spart})--(\ref{Tpart}), having the structure of $\mathcal{J}_L$
and $\mathcal{K}$, are computed by surface integrals at infinity using
the properties of the analytic continuation in $B$. The only task is
to look for the relevant coefficients in the expansions of the
integrands at infinity, (\ref{Gexp}) or (\ref{Hiexp}), and to perform
the surface integrals (\ref{Jfinal}) or (\ref{Kfinal}). This saves a
lot of calculations with respect to the ``bulk'' calculation of the
Hadamard partie finie based on the form found in Eq. (\ref{Ires}). Of
course, the two calculations, at infinity and in the bulk, will
completely agree, but notice that for this agreement to work, one must
crucially take into account in the bulk calculation, in addition to
the formula (\ref{Ires}), the contribution of the distributional part
of derivatives. Thus the Laplacian in Eq. (\ref{J}) is to be
considered in a distributional sense.

\section{Multipole moments of point particle binaries}\label{pointpart}

\subsection{The 3PN mass quadrupole moment}\label{MQ}

We have computed the 3PN mass quadrupole moment of point particle
binaries, for general orbits, using the expressions
(\ref{Spart})--(\ref{Tpart}) with $\ell=2$, following the rules of the
pHS regularization, notably the way (\ref{calFdelta}) one handles the
compact-support ``contact'' terms, and the techniques reviewed in
Sections \ref{HadPf} and \ref{infbound} to compute three-dimensional
non-compact-support integrals. We denote by
$I^\mathrm{\,pHS}_{ij}[s_1,s_2,r_0]$ the result of such pHS
calculation, in which $s_1$ and $s_2$ denote the two UV cut-offs and
$r_0$ the IR length scale. These constants result from the computation
of non-compact-support integrals and are shown in our basic formula
(\ref{Ires}) for the Hadamard partie-finie integral. Now it was argued
in paper I that the Hadamard regularization of the 3PN quadrupole
moment is incomplete, and must be augmented, in order not to be
incorrect, by some unknown, ambiguous, contributions.

The first source of ambiguity is the ``kinetic'' one, linked to the
inability of the Hadamard regularization to ensure the global
Poincar\'e invariance of the formalism (we are speaking here of the
ordinary or pHS variants of the Hadamard regularization, as well as of
the ``hybrid'' regularization which has been used in paper I for the
generation problem\,\footnote{An exception is the extended Hadamard
regularization which is in principle able to preserve the Lorentz
invariance, since the Hadamard regularization is performed in the
Lorentzian rest frame of each of the particles \cite{BFregM}. However
we have not been able to fix the kinetic ambiguity in the 3PN
quadrupole moment using the extended Hadamard regularization.}). As
discussed in Section X of paper I we must account for the kinetic
ambiguity by adding ``by hands'' a specific ambiguity term, depending
on a single ambiguity parameter called $\zeta$. Following here exactly
the same reasoning we add to the pHS result the same type of ambiguous
term, which means that we must consider as correct the following 3PN
quadrupole moment,
\begin{eqnarray}\label{Iij}
I_{ij}[s_1,s_2,r_0;\hat{\zeta}] &=&
I^\mathrm{\,pHS}_{ij}[s_1,s_2,r_0]\nonumber\\&+&\frac{44\,\hat{\zeta}}{3}
\,\frac{G^2\,m_1^3}{c^6}\,v_1^{<i}v_1^{j>}+1\leftrightarrow 2\,,
\end{eqnarray}
where the extra term, purely of 3PN order, involves an unknown
coefficient $\hat{\zeta}$. Here the coordinate velocity is denoted
$v_1^i$, and the factor $44/3$ is chosen for convenience. The
parameter $\hat{\zeta}$ will turn out to be different from the
parameter $\zeta$ of paper I because we are adding it to the result of
the pHS regularization, instead of the ``hybrid'' Hadamard-type
regularization considered in paper I. [The hybrid regularization
differs from the pHS one by the way the contact terms are computed,
which takes into account the properties of non-distributivity
(\ref{nondistr}) and (\ref{deltanondistr}) and is more like the one of
the extended Hadamard regularization \cite{BFreg}, and in some subtle
differences arising between the ``case-by-case'' computation of the
elementary non-compact integrals in paper I and the systematic
approach followed here which is based on the formula (\ref{Ires}). Of
course, there is only one thing which is finally important, namely
that these differences are completely encoded into some mere shifts of
the values of the ambiguity parameters, see Eqs. (\ref{hatamb})
below.]

The second source of ambiguity is ``static''. It comes from the
\textit{a priori} unknown relation between the Hadamard regularization
length scales, $s_1$ and $s_2$, and the ones, $r'_1$ and $r'_2$,
parametrizing the 3PN equations of motion in harmonic coordinates
\cite{BF00,BFeom}. The constants $r'_1$ and $r'_2$ come from the
regularization of Poisson-type integrals in the computation of the
equations of motion, and can be interpreted as some infinitesimal
radial distances used as cut-offs when the field point tends to the
singularities. Since we need the equations of motion when computing
the time derivatives of the 3PN quadrupole moment, for instance in
order to obtain the gravitational-wave flux, we must definitely know
the relation between $s_1$, $s_2$ and $r'_1$, $r'_2$. This relation
constitutes a true physical undeterminacy within the various variants
of Hadamard's regularization (either ordinary, pHS, hybrid or
extended). Let us rewrite the R.H.S. of Eq. (\ref{Iij}) by
``artificially'' introducing $r'_1$ and $r'_2$ into the two slots of
the pHS result. For doing this we use the known dependence of the pHS
quadrupole in terms of the constants $s_1$ and $s_2$. This dependence
is the same as in the case of the ``hybrid'' quadrupole (and indeed of
any other of its regularization variants), and is given by Eq. (10.4)
of paper I. Hence we have
\begin{equation}\label{IpHSlog} 
I^\mathrm{\,pHS}_{ij}[s_1,s_2,r_0] =
\frac{44}{3}\,\frac{G^2\,m_1^3}{c^6}\,\ln\left(
\frac{r_{12}}{s_1}\right)\,y_1^{<i}a_1^{j>}+1\leftrightarrow
2+\cdots\,,
\end{equation}
where $a_1^i$ denotes the Newtonian acceleration and the dots indicate
the terms that are independent of $s_1$ and $s_2$ (but which can
depend on $r_0$). Using this structure it is evident that the effect
of changing $s_1,s_2\rightarrow r'_1,r'_2$ in the pHS quadrupole is
\begin{eqnarray}\label{changeIpHSlog} 
I^\mathrm{\,pHS}_{ij}[s_1,s_2,r_0] &=&
I^\mathrm{\,pHS}_{ij}[r'_1,r'_2,r_0] \nonumber\\&+&
\frac{44}{3}\,\frac{G^2\,m_1^3}{c^6}\,\ln\left(
\frac{r'_1}{s_1}\right)\,y_1^{<i}a_1^{j>}+1\leftrightarrow 2\,.
\end{eqnarray}
We now argue, exactly like in Section X of paper I, that the most
general admissible structure for the unknown logarithmic ratio in
Eq. (\ref{changeIpHSlog}) is
\begin{equation}\label{log}
\ln\left( \frac{r'_1}{s_1}\right) = \hat{\xi} +
\hat{\kappa}\,\frac{m_1+m_2}{m_1}~~\hbox{and $1\leftrightarrow 2$}\,,
\end{equation}
where $\hat{\xi}$ and $\hat{\kappa}$ denote two new arbitrary
ambiguity parameters, which are also \textit{a priori} different from
$\xi$ and $\kappa$ in paper I. The argument leading to Eq. (\ref{log})
is essentially that the quadrupole moment should be a polynomial in
the two masses $m_1$ and $m_2$ separately. Therefore,
\begin{eqnarray}\label{changeIpHSlog'} 
I^\mathrm{\,pHS}_{ij}[s_1,s_2,r_0] &=&
I^\mathrm{\,pHS}_{ij}[r'_1,r'_2,r_0] \nonumber\\&+&
\frac{44}{3}\,\frac{G^2\,m_1^3}{c^6}\,\left(\hat{\xi} +
\hat{\kappa}\,\frac{m_1+m_2}{m_1}\right)\,y_1^{<i}a_1^{j>}+1\leftrightarrow
2\,,
\end{eqnarray}
so we write the Hadamard-regularized 3PN quadrupole, depending on the
three ambiguity parameters $\hat{\xi}$, $\hat{\kappa}$ and
$\hat{\zeta}$, in the form
\begin{eqnarray}\label{Iijres}
I_{ij}[s_1,s_2,r_0;\hat{\zeta}] &=&
I^\mathrm{\,pHS}_{ij}[r'_1,r'_2,r_0]\nonumber\\&+&\frac{44}{3}
\,\frac{G^2\,m_1^3}{c^6}\,\left[\left(\hat{\xi} +
\hat{\kappa}\,\frac{m_1+m_2}{m_1}\right)\,y_1^{<i}a_1^{j>}
+\hat{\zeta}\,v_1^{<i}v_1^{j>}\right] +1\leftrightarrow 2\,.~~~
\end{eqnarray}
We recall from paper I that, by contrast with the latter ambiguity
parameters, the three scales $r'_1$, $r'_2$ and $r_0$ are not physical
and must disappear from the final results (when they are expressed in
a coordinate-invariant way).

Next let us compare the result, for \textit{general non-circular}
orbits, with the one of paper I which was obtained by means of the
hybrid Hadamard-type regularization.\,\footnote{For the purpose of the
comparison we have redone the calculation of paper I in the case of
non-circular orbits. In fact, although the end result of paper I is
presented for circular orbits, most of the intermediate expressions in
this paper are valid for general binary orbits.} If everything is
consistent, Eq. (\ref{Iijres}) should be in perfect agreement with
paper I modulo a change of definition of the three ambiguity
parameters, due to the use of the pHS regularization here instead of
the hybrid regularization in paper I. We find that indeed there is a
complete match for all the terms with those of paper I if and only if
the ambiguity parameters $\hat{\xi}$, $\hat{\kappa}$ and $\hat{\zeta}$
are related to the corresponding ones $\xi$, $\kappa$ and $\zeta$ in
paper I by
\begin{subequations}\label{hatamb}\begin{eqnarray}
\hat{\xi} &=& \xi + \frac{1}{22}\,,\\ \hat{\kappa} &=& \kappa\,,\\
\hat{\zeta} &=& \zeta + \frac{9}{110}\,.
\end{eqnarray}\end{subequations}
In view of the many differences between the present calculation and
the one of paper I (\textit{e.g.} in the definition of the
regularization, the choice of elementary potentials, the way one
computes non-compact support integrals), this agreement constitutes an
important check of the lengthy algebra and the correctness of the
result. In the following we prefer to come back to the original
ambiguity parameters $\xi$, $\kappa$ and $\zeta$ adopted in paper I,
so we write the quadrupole moment as\,\footnote{We employ the slightly
abusive notation that $I_{ij}[s_1,s_2,r_0;\hat{\zeta}]\equiv
I_{ij}[r'_1,r'_2,r_0;\xi,\kappa,\zeta]$ when Eqs. (\ref{log}) and
(\ref{hatamb}) hold.}
\begin{eqnarray}\label{Iijres'}
I_{ij}[r'_1,r'_2,r_0;\xi,\kappa,\zeta] &=&
I^\mathrm{\,pHS}_{ij}[r'_1,r'_2,r_0]\nonumber\\&+&\frac{44}{3}
\,\frac{G^2\,m_1^3}{c^6}\,\left[\left(\xi + \frac{1}{22}+
\kappa\,\frac{m_1+m_2}{m_1}\right)\,y_1^{<i}a_1^{j>} +\left(\zeta +
\frac{9}{110}\right)\,v_1^{<i}v_1^{j>}\right]
\nonumber\\&+&1\leftrightarrow 2\,.
\end{eqnarray}

Finally we present the result of the computation of all the terms in
Eqs. (\ref{Spart})--(\ref{Tpart}) for $\ell=2$ and general binary
orbits. Unfortunately we find that the end expression of the
quadrupole is very long in a general frame (with arbitrary origin), so
we decide to present only the much shorter expression valid in the
frame of the center of mass. The center-of-mass frame is defined by
the nullity of the 3PN conserved integral of the center-of-mass vector
deduced from the 3PN equations of motion in harmonic coordinates
\cite{ABF01}. For this calculation we use the relations between the
general and center-of-mass frames given at 3PN order in
Ref. \cite{BI03CM}. The structure of the 3PN center-of-mass quadrupole
moment is\,\footnote{Our notation is $m\equiv m_1+m_2$ and $\nu\equiv
m_1m_2/m^2$; $x^i\equiv y_1^i-y_2^i$ and $v^i\equiv dx^i/dt =
v_1^i-v_2^i$; $v^2=\mathbf{v}^2$ and $\dot{r}=\mathbf{n}.\mathbf{v}$,
where $\mathbf{n}\equiv\mathbf{x}/r$ and
$r\equiv\vert\mathbf{x}\vert$; the STF projection is indicated by
brackets surrounding the indices.}
\begin{eqnarray}
\label{Iijfinal}
I_{ij}[r'_1,r'_2,r_0;\xi,\kappa,\zeta]&=&\nu\,m
\,\biggl\{\left[\mathcal{A}-\frac{24}{7}\,\frac{\nu}{c^5}\,\frac{G^2
m^2}{r^2}\,\dot{r}\right]\, x_{\langle
i}x_{j\rangle}+\mathcal{B}\,\frac{r^2}{c^2}\,v_{\langle i}v_{j\rangle}
\nonumber\\&&\quad\quad +
2\left[\mathcal{C}\,\frac{r\,\dot{r}}{c^2}+\frac{24}{7}
\,\frac{\nu}{c^5}\,\frac{G^2 m^2}{r}\right]\,x_{\langle i}v_{j\rangle}
\biggr\}\,.
\end{eqnarray}
Here we have explicitly displayed the ``odd'' 2.5PN radiation reaction
contributions. The content of the ``even'' terms is given by the
coefficients $\mathcal{A}$, $\mathcal{B}$ and $\mathcal{C}$, which
generalize to non-circular orbits those given in Eqs. (11.3)--(11.4)
of paper I, and read
{\allowdisplaybreaks\begin{subequations}\label{ABC}\begin{eqnarray}
\mathcal{A}&=& 1
+\frac{1}{c^2} \left[v^2\, \left( \frac{29}{42}
  - \frac{29\,\nu }{14} \right)+\frac{G\,m}{r}\, \left( -\frac{5}{7}+
  \frac{8 }{7}\,\nu \ \right) \right]\nonumber\\
&&+ \frac{1}{c^4}\left[v^2\,\frac{G\,m}{r}\,\left( \frac{2021}{756} -
\frac{5947 }{756}\,\nu-\frac{4883}{756}\,\nu^2
\right)\right.\nonumber\\&&\qquad\left.+\left(\frac{G\,m}{r}\right)^2\,\left(
- \frac{355}{252} -\frac{953 }{126}\,\nu + \frac{337\,}{252}\,\nu^2
\right)\right.\nonumber\\ &&\qquad+\left.v^4\,\left( \frac{253}{504} -
\frac{1835 }{504}\,\nu
+\frac{3545}{504}\,\nu^2\right)\right.\nonumber\\&&\qquad
\left.\left. +\dot{r}^2\, \frac{G\,m}{r}\,\left( - \frac{131}{756} +
\frac{907 }{756}\,\nu - \frac{1273}{756}\,\nu^2
\right)\right)\right]\nonumber\\
&&+\frac{1}{c^6}\left[v^6\,\left( \frac{4561}{11088} -
\frac{7993}{1584}\,\nu+\frac{117067}{5544}\,\nu^2 -
\frac{328663}{11088}\,\nu^3\right)\right.\nonumber\\
&&\left.\qquad+v^4\, \frac{G\,m}{r}\,\left( \frac{307}{77} - \frac{94475
}{4158}\,\nu+\frac{218411}{8316}\,\nu^2 +
\frac{299857}{8316}\,\nu^3\right)\right.\nonumber\\
&&\qquad+\left(\frac{G\,m}{r}\right)^3\,\left(
\frac{6285233}{207900}+\frac{34091}{1386}\,\nu
-\frac{3632}{693}\,\nu^2 + \frac{13289}{8316}\,\nu^3
\right.\nonumber\\
&&\qquad\quad\left. -\frac{44}{3}\,\nu\,\left(\xi+2\kappa\right)
-\frac{428}{105}\,\ln \left(\frac{r}{r_0}\right) - \frac{44}{3}\,\,\nu
\,\ln \left(\frac{r}{r_0'}\right) \right)\nonumber\\
&&\qquad+{\dot{r}}^2\,\left(\frac{G\,m}{r}\right)^2\,\left( -
\frac{8539}{20790}+ \frac{52153 }{4158}\,\nu - \frac{4652}{231}\,\nu^2
-\frac{54121}{5544}\,\nu^3 \right) \,\nonumber\\
&&\qquad+{\dot{r}}^4\,\frac{G\,m}{r}\,\left( \frac{2}{99} -
\frac{1745}{2772}\,\nu +\frac{16319}{5544}\,\nu^2 -
\frac{311\,}{99}\,\nu^3 \right) \,\nonumber\\
&&\qquad+v^2\,\left(\frac{G\,m}{r}\right)^2\,\left( \frac{187183}{83160}
-\frac{605419 }{16632}\,\nu + \frac{434909}{16632}\,\nu^2
-\frac{37369}{2772}\,\nu^3 \right)\nonumber\\
&&\qquad+\left.v^2\,\frac{G\,m}{r}\,\,{\dot{r}}^2 \,\left( -
\frac{757}{5544}+\frac{5545 }{8316}\,\nu -
\frac{98311\,}{16632}\,\nu^2 +\frac{153407}{8316}\,\nu^3 \right)
\right]\,, \\
\mathcal{B}&=& \frac{11}{21} - \frac{11}{7}\,\nu\nonumber\\ &&
+\frac{1}{c^2}\left[\frac{G m}{r}\,\left( \frac{106}{27} -
\frac{335}{189}\,\nu - \frac{985}{189}\,\nu^2
\right)\right. \nonumber\\ &&\qquad\left. +\,v^2\,\left(
\frac{41}{126} - \frac{337 }{126}\,\nu + \frac{733}{126}\,\nu^2
\right)+\dot{r}^2\,\left( \frac{5\,}{63} - \frac{25 }{63}\,\nu +
\frac{25}{63}\,\nu^2 \right)\right] \nonumber\\ &&
+\frac{1}{c^4}\,\left[\,v^4\,\left( \frac{1369}{5544} - \frac{19351
}{5544}\,\nu + \frac{45421}{2772}\,\nu^2 - \frac{139999}{5544}\,\nu^3
\right) \right.\nonumber\\
&&\qquad +\left.\left(\frac{G
m}{r}\right)^2\,\left(-\frac{40716}{1925}- \frac{4294 }{2079}\,\nu
+\frac{62576}{2079}\,\nu^2 - \frac{24314}{2079}\,\nu^3
\right.\right.\nonumber\\ &&\qquad\quad\left.\left.  +
\frac{428}{105}\,\ln
\left(\frac{r}{r_0}\right)+\frac{44}{3}\,\nu\,\zeta
\right)\right.\nonumber\\ &&\qquad+\dot{r}^2\,\frac{G m}{r}\left(
\frac{79}{77}-\frac{5807}{1386}\,\nu + \frac{515}{1386}\,\nu^2
+\frac{8245}{693}\,\nu^3 \right)\nonumber\\ &&\qquad+ v^2\,\frac{G
m}{r}\,\left( \frac{587}{154} - \frac{67933 }{4158}\,\nu+
\frac{25660}{2079}\,\nu^2 +\frac{129781}{4158}\,\nu^3
\right)\nonumber\\ &&\qquad+\left. v^2\,\dot{r}^2\,\left(
\frac{115\,}{1386}-\frac{1135}{1386}\,\nu
+\frac{1795}{693}\,\nu^2-\frac{3445}{1386}\,\nu^3 \right)\right]\,,\\
\mathcal{C}&=& -\frac{2}{7}+ \frac{6}{7}\,\nu \nonumber\\ &&+
\frac{1}{c^2}\left[v^2\, \left( -\frac{13}{63} + \frac{101 }{63}\,\nu
- \frac{209\,}{63}\,\nu^2 \right) \right.\nonumber\\ &&\qquad
\left.+\frac{G m}{r}\,\left( -\frac{155}{108} + \frac{4057}{756}\,\nu
+ \frac{209}{108}\,\nu^2 \right)\right] \nonumber\\
&&+\frac{1}{c^4}\left[v^2\,\frac{G m}{r}\, \left( - \frac{2839}{1386}+
\frac{237893}{16632}\,\nu - \frac{188063}{8316}\,\nu^2
-\frac{58565}{4158}\,\nu^3 \right)\right.\nonumber\\
&&\qquad+\,\left(\frac{G m}{r}\right)^2\,\left( -\frac{12587}{41580}+
\frac{406333 }{16632}\,\nu - \frac{2713}{396}\,\nu^2
+\frac{4441}{2772}\,\nu^3 \right)\nonumber\\ &&\qquad+v^4\,\left( -
\frac{457}{2772}+ \frac{6103 }{2772}\,\nu -
\frac{13693}{1386}\,\nu^2+\frac{40687}{2772}\,\nu^3 \right)\nonumber\\
&&\qquad+\left. {\dot{r}^2\,\frac{G m}{r}\,\left( \frac{305}{5544} +
\frac{3233}{5544}\,\nu - \frac{8611}{5544}\,\nu^2 -
\frac{895}{154}\,\nu^3\right) }\right]\,.
\end{eqnarray}\end{subequations}}\noindent
The 3PN quadrupole moment depends on $\xi$, $\kappa$ and $\zeta$, on
the constant scale $r_0$ introduced into the general formalism defined
for extended PN sources in Eq. (\ref{rtilde}), and on the
``logarithmic barycenter'' $r'_0$ of the two Hadamard self-field
regularization scales $r'_1$ and $r'_2$, defined by
\begin{equation}\label{r'0}
m\,\ln r'_0 = m_1\,\ln r'_1 + m_2\,\ln r'_2\,.
\end{equation}
Unlike $r_0$ which cancels out in the complete waveform, already at
the level of the general ``fluid'' formalism, and $r'_0$ which
represents some gauge constant devoid of physical meaning (see
\cite{BFeom} and paper I), the ambiguity parameters $\xi$, $\kappa$
and $\zeta$ represent some genuine physical unknowns, which have
recently been computed by means of dimensional regularization in
Ref. \cite{BDEI04}. We shall show in the next Section that it is
possible to determine a particular combination of these parameters in
the context of Hadamard's regularization.

\subsection{The 3PN mass dipole moment}\label{MD}

Recall from Section \ref{monodipo} that the mass-type dipole moment or
``ADM dipole moment'' $M_i$, which varies linearly with time
($\ddot{M}_i=0$), is the sum of two terms,
\begin{equation}\label{Mi}
M_i = I_i + \delta I_i\,,
\end{equation}
where $I_i$ is defined by the same general expression (\ref{IL}) as
for non-conserved moments but in which we set $\ell=1$, and where the
suplementary piece $\delta I_i$ is given by Eq. (\ref{DeltaIicons}).

We first concentrate our attention on the first part $I_i$ which is
thus given, up to 3PN order, by the explicit expressions
(\ref{Spart})--(\ref{Tpart}) with $\ell=1$. We follow the same steps
as for the quadrupole moment investigated in Section
\ref{MQ}. Repeating the arguments presented in Section X of paper I,
we notice first that in the case of the dipole moment there is no
ambiguous terms of the ``kinetic'' type. Actually one can easily check
on dimensional grounds that the existence of such a term in $I_i$,
which would be proportional to $v_1^i$ (plus $1\leftrightarrow 2$), is
impossible. Thus, unlike in the quadrupole case as shown in
(\ref{Iij}), $I_i$ is directly given by the result of the pHS
regularization, 
\begin{equation}\label{Ii}
I_i[s_1,s_2] = I^\mathrm{\,pHS}_i[s_1,s_2]\,.
\end{equation}
Here $s_1$ and $s_2$ are the two regularization scales coming from
Eq. (\ref{Pf}), but as it turns out that there is no dependence on the
cut-off scale $r_0$ in the dipolar case. To define the static
ambiguity we must now re-express the dipole moment in terms of the
particular equation-of-motion-related scales $r'_1$ and $r'_2$. For
this we use the dependence of the dipole moment in terms of the scales
$s_1,s_2$,
\begin{equation}\label{IipHSlog} 
I^\mathrm{\,pHS}_i[s_1,s_2] =
\frac{22}{3}\,\frac{G^2\,m_1^3}{c^6}\,a_1^i\ln\left(
\frac{r_{12}}{s_1}\right)+1\leftrightarrow 2+\cdots\,,
\end{equation}
where the dots represent the terms that are independent of $s_1$ and
$s_2$. Notice the factor $22/3$ instead of $44/3$ in the quadrupolar
case (\ref{IpHSlog}). This yields immediately
\begin{eqnarray}\label{Iilog}
I^\mathrm{\,pHS}_i[s_1,s_2] &=&
I^\mathrm{\,pHS}_i[r'_1,r'_2]\nonumber\\&+&\frac{22}{3}
\,\frac{G^2\,m_1^3}{c^6}\,a_1^i\,\ln\left(
\frac{r'_1}{s_1}\right)+1\leftrightarrow 2\,.
\end{eqnarray}
The ratio $r'_1/s_1$ is \textit{a priori} unknown but we remember that
it has already served for the definition of two of our ambiguity
parameters: $\hat{\xi}$ and $\hat{\kappa}$, see Eq. (\ref{log}). Now
we shall use in our present calculation of the dipole moment the
\textit{same relation} between $r'_1$ and $s_1$ as was used for the
quadrupole moment. This means that we consider that the constants
$s_1$ and $s_2$ parametrizing the Hadamard partie finie (\ref{Pf})
have been chosen once and for all at the beginning of both our
calculations of the quadrupole and dipole moments, where they take
some definite meaning related for instance to the shape of the
regularizing volumes $B_1$ and $B_2$ which are initially excised
around the two singularities when applying Hadamard's definition in
the form of Eq. (\ref{Pf}). Thus we assume that $s_1$ and $s_2$
represent some unknown but \textit{fixed} constants --- having the
same values for the two calculations of the quadrupole and dipole
moments.\,\footnote{We tried to further extend this type of argument
to the constants $s_1$ and $s_2$ which were used in the 3PN equations
of motion \cite{BF00,BFeom}. This implied that we had to use for the
wave generation the same regularization as in the equations of motion,
\textit{i.e.} the extended Hadamard regularization
\cite{BFreg,BFregM}. Unfortunately this program, whose aim would have
been to determine all the ambiguity parameters within Hadamard's
regularization ($\xi$, $\kappa$, $\zeta$ and also $\lambda$), did not
fully succeed. Nevertheless a less ambitious part of the program did
succeed, and this is what we show here.} When substituting the
expression $\ln\left( r'_1/s_1\right) = \hat{\xi} + \hat{\kappa}+
\hat{\kappa}\,m_2/m_1$ into Eq. (\ref{Iilog}) we observe that the last
term, which is proportional to the mass ratio $m_2/m_1$, cancels out
after applying the symmetry exchange $1\leftrightarrow 2$. So we find
that the dipole moment depends in fact on \textit{one and only one}
combination of ambiguity parameters, namely $\hat{\xi}+\hat{\kappa}$,
and we obtain
\begin{eqnarray}\label{Iires}
I^\mathrm{\,pHS}_i[s_1,s_2] &=&
I^\mathrm{\,pHS}_i[r'_1,r'_2]\nonumber\\&+&\frac{22}{3}
\,\frac{G^2\,m_1^3}{c^6}\left(\hat{\xi} + \hat{\kappa}
\right)\,a_1^i+1\leftrightarrow 2\,.
\end{eqnarray}
Then we come back to the original definitions of paper I by using
Eqs. (\ref{hatamb}), and this leads to the following expression of the
3PN dipole moment:
\begin{eqnarray}\label{Iires'}
I_i[r'_1,r'_2;\xi+\kappa] &=&
I^\mathrm{\,pHS}_i[r'_1,r'_2]\nonumber\\&+&\frac{22}{3}
\,\frac{G^2\,m_1^3}{c^6}\left(\xi + \kappa + \frac{1}{22}
\right)\,a_1^i+1\leftrightarrow 2\,.
\end{eqnarray}

At this stage we have to worry about the extra contribution $\delta
I_i$ present in Eq. (\ref{Mi}). From its expression given by
(\ref{DeltaIicons}) we see that obtaining $\delta I_i$ at the 3PN
order requires both $\Sigma_a$ and $\Sigma_{ab}$ with the full 3PN
accuracy. By contrast, recall that the calculation of $I_i$
necessitated $\Sigma$ at the 3PN order, but $\Sigma_a$ and
$\Sigma_{ab}$ with only the 2PN and 1PN precisions respectively. Thus
it seems that $\delta I_i$ cannot be obtained solely with the formulas
developed in Section \ref{3PNmoment}. Notice that the expression of
$\delta I_i$ involves an explicit factor $B$, and thus depends only on
the presence of IR poles $\propto 1/B$ in the integrals. Consequently
$\delta I_i$ can be computed by the same techniques as in Section
\ref{infbound}, \textit{i.e.} in the form of surface integrals at
infinity similar to Eqs. (\ref{Jfinal}) or (\ref{Kfinal}). We have
been able to prove that all the terms in $\delta I_i$ are
\textit{separately zero} up to the 3PN order. For all the terms we did
know from using the results of Section \ref{3PNmoment} we have made a
complete calculation, and for the other terms we looked at their
allowed \textit{structure} in terms of the basic potentials $V$,
$V_i$, $\hat{W}_{ij}$, $\cdots$, invoking dimensionality arguments but
leaving aside the unimportant numerical coefficients, which was
sufficient to check that the corresponding surface integrals are
exactly zero for all the terms. Thus, we conclude that $\delta I_i=0$
at 3PN order, hence
\begin{equation}\label{DeltaIzero}
M_i = I_i + \mathcal{O}\left(c^{-7}\right)\,,
\end{equation}
which finally results, from the detailed evaluation of all the terms
in Eqs. (\ref{Spart})--(\ref{Tpart}), in\,\footnote{The two masses
$m_1$ and $m_2$ are located at the positions $\mathbf{y}_1$ and
$\mathbf{y}_2$, the unit vector between them is
$\mathbf{n}_{12}=(\mathbf{y}_1-\mathbf{y}_2)/r_{12}$ with
$r_{12}=\vert\mathbf{y}_1-\mathbf{y}_2\vert$, the two coordinate
velocities are $\mathbf{v}_1=d\mathbf{y}_1/dt$ and
$\mathbf{v}_2=d\mathbf{y}_2/dt$, and
$\mathbf{v}_{12}=\mathbf{v}_1-\mathbf{v}_2$. Euclidean scalar products
are denoted by parenthesis, \textit{e.g.}
$(n_{12}v_1)=\mathbf{n}_{12}.\mathbf{v}_1$.}  {\allowdisplaybreaks
\begin{eqnarray}\label{Mifinal}
M_i &=& m_1 \,y_1^i \nonumber \\ &+& \frac{1}{c^2}
\biggl\{y_1^i\bigg(-\frac{G \,m_1 \,m_2}{2 r_{12}} + \frac{ m_1
\,v_1^2}{2}\bigg) \biggr\} \nonumber \\ &+& \frac{1}{c^4}\biggl\{ G
\,m_1 \,m_2 \,v_1^i\bigg(-\frac{7}{4} (n_{12}v_1) - \frac{7}{4}
(n_{12}v_2)\bigg) \nonumber\\ && \quad + y_1^i
\,\bigg(-\frac{5}{4}\frac{G^2 \,m_1^2 \,m_2}{r_{12}^2} +
\frac{7}{4}\frac{G^2 \,m_1 \,m_2^2}{r_{12}^2} + \frac{3}{8} m_1\,v_1^4
\nonumber \\ & & \quad\quad + \frac{G \,m_1 \,m_2}{r_{12}}
\left[-\frac{1}{8} (n_{12}v_1)^2 - \frac{1}{4} (n_{12}v_1) (n_{12}v_2)
+ \frac{1}{8} (n_{12}v_2)^2 \right.\nonumber \\ & &
\left.\quad\quad\quad + \frac{19}{8} v_1^2 - \frac{7}{4} (v_1v_2) -
\frac{7}{8} v_2^2\right]\bigg)\biggr\} \nonumber \\ &+& \frac{1}{c^6}
\biggl\{ v_1^i \bigg( \frac{235}{24}\frac{G^2 \,m_1^2 \,m_2}{r_{12}}
(n_{12}v_{12}) - \frac{235}{24}\frac{G^2 \,m_1 \,m_2^2}{r_{12}}
(n_{12}v_{12}) \nonumber \\ & & \quad\quad + G \,m_1 \,m_2
\left[\frac{5}{12} (n_{12}v_1)^3 + \frac{3}{8} (n_{12}v_1)^2
(n_{12}v_2) + \frac{3}{8} (n_{12}v_1) (n_{12}v_2)^2 \right.\nonumber
\\ & & \left.\quad\quad\quad + \frac{5}{12} (n_{12}v_2)^3 -
\frac{15}{8} (n_{12}v_1) v_1^2 - (n_{12}v_2) v_1^2 + \frac{1}{4}
(n_{12}v_1) (v_1v_2) \right.\nonumber \\ & & \left.\quad\quad\quad +
\frac{1}{4} (n_{12}v_2) (v_1v_2) - (n_{12}v_1) v_2^2 - \frac{15}{8}
(n_{12}v_2) v_2^2\right]\bigg) \nonumber \\ & & \quad + y_1^i \,\bigg(
\frac{5}{16} m_1 \,v_1^6 \nonumber \\ & & \quad\quad + \frac{G \,m_1
\,m_2}{r_{12}} \left[\frac{1}{16} (n_{12}v_1)^4 + \frac{1}{8}
(n_{12}v_1)^3 (n_{12}v_2) + \frac{3}{16} (n_{12}v_1)^2 (n_{12}v_2)^2
\right.\nonumber \\ & & \left.\quad\quad\quad + \frac{1}{4}
(n_{12}v_1) (n_{12}v_2)^3 - \frac{1}{16} (n_{12}v_2)^4 - \frac{5}{16}
(n_{12}v_1)^2 v_1^2 \right.\nonumber \\ & & \left.\quad\quad\quad -
\frac{1}{2} (n_{12}v_1) (n_{12}v_2) v_1^2 - \frac{11}{8} (n_{12}v_2)^2
v_1^2 + \frac{53}{16} v_1^4 + \frac{3}{8} (n_{12}v_1)^2 (v_1v_2)
\right.\nonumber \\ & & \left.\quad\quad\quad + \frac{3}{4}
(n_{12}v_1) (n_{12}v_2) (v_1v_2) + \frac{5}{4} (n_{12}v_2)^2 (v_1v_2)
- 5 v_1^2 (v_1v_2) \right.\nonumber \\ & & \left.\quad\quad\quad +
\frac{17}{8} (v_1v_2)^2 - \frac{1}{4} (n_{12}v_1)^2 v_2^2 -
\frac{5}{8} (n_{12}v_1) (n_{12}v_2) v_2^2 + \frac{5}{16} (n_{12}v_2)^2
v_2^2 \right.\nonumber \\ & & \left.\quad\quad\quad + \frac{31}{16}
v_1^2 v_2^2 - \frac{15}{8} (v_1v_2) v_2^2 - \frac{11}{16}
v_2^4\right]\nonumber \\ & & \quad\quad + \frac{G^2 \,m_1^2
\,m_2}{r_{12}^2} \left[\frac{79}{12} (n_{12}v_1)^2 - \frac{17}{3}
(n_{12}v_1) (n_{12}v_2) \right.\nonumber \\ & & \left.\quad\quad\quad
+ \frac{17}{6} (n_{12}v_2)^2 - \frac{175}{24} v_1^2 + \frac{40}{3}
(v_1v_2) - \frac{20}{3} v_2^2\right]\nonumber \\ & & \quad\quad +
\frac{G^2 \,m_1 \,m_2^2}{r_{12}^2} \left[-\frac{7}{3} (n_{12}v_1)^2 +
\frac{29}{12} (n_{12}v_1) (n_{12}v_2) + \frac{2}{3} (n_{12}v_2)^2
\right.\nonumber \\ & & \left.\quad\quad\quad + \frac{101}{12} v_1^2 -
\frac{40}{3} (v_1v_2) + \frac{139}{24} v_2^2 \right]\nonumber \\ & &
\quad\quad -\frac{19}{8}\frac{G^3 \,m_1^2 \,m_2^2}{r_{12}^3}\nonumber
\\ & & \quad\quad + \frac{G^3 \,m_1^3 \,m_2}{r_{12}^3}
\left[\frac{55}{18} - \frac{22}{3}(\xi + \kappa) - \frac{22}{3} \ln
\left(\frac{r_{12}}{r'_1} \right)\right] \nonumber \\ & & \quad\quad +
\frac{G^3 \,m_1 \,m_2^3}{r_{12}^3} \left[-\frac{32}{9} +
\frac{22}{3}(\xi + \kappa) + \frac{22}{3} \ln
\left(\frac{r_{12}}{r'_2} \right)\right]\bigg) \biggr\} + 1
\leftrightarrow 2\,.
\end{eqnarray}}
 
The case of the conserved dipole moment is interesting because it
offers us a very good check of the calculations. Indeed let us argue
that $M_i$, which represents the distribution of positions of
particles as weighted by their \textit{gravitational masses} $m_g$,
must be \textit{identical} to the position of the center of mass $G_i$
of the system of particles (per unit of total mass), because the
center of mass $G_i$ represents in fact the same quantity as the
dipole $I_i$ but corresponding to the \textit{inertial masses} $m_i$
of the particles. The equality between mass dipole $M_i$ and
center-of-mass position $G_i$ can thus be seen as a consequence of the
equivalence principle $m_i=m_g$, which is surely incorporated in our
model of point particles. Now the center of mass $G_i$ is already
known at the 3PN order for point particle binaries, as one of the
conserved integrals of the 3PN motion in harmonic coordinates (we
neglect the radiation-reaction term at 2.5PN order). We recall that
$G_i$, given explicitly in Ref. \cite{ABF01}, depends on the
regularization length scales $r'_1$ and $r'_2$ which are \textit{the
same} as in the 3PN equations of motion \cite{BF00,BFeom} and
therefore the same as in our result (\ref{Mifinal}) --- by
\textit{definition} of the ambiguity parameters $\xi$ and
$\kappa$. However $G_i$ was found in Ref. \cite{ABF01} to be free of
ambiguities; for instance the ambiguity parameter $\lambda$ in the 3PN
equations of motion disappears from the expression of $G_i$. Let us
therefore impose the equivalence between $M_i$ and $G_i$, which means
we make the complete identification
\begin{equation}\label{MiGi}
M_i[r'_1,r'_2;\xi+\kappa] \equiv G_i[r'_1,r'_2]\,,
\end{equation}
in which we insist that the constants $r'_1$ and $r'_2$ appearing in
both sides of this equation are the same. Comparing $M_i$ with the
expression of $G_i$ given by Eq. (4.5) in \cite{ABF01}, we find that
these constants $r'_1$ and $r'_2$ cancel out, and that
Eq. (\ref{MiGi}) is verified for all the terms \textit{if and only if}
the particular combination of ambiguity parameters $\xi+\kappa$ is
fixed to the unique value
\begin{equation}\label{xikappa}
\xi + \kappa = -\frac{9871}{9240}\,.
\end{equation}
This result is obtained within Hadamard's regularization. It shows
that, although as we have seen Hadamard's regularization is
``physically incomplete'' (at 3PN order), it can nevertheless be
partially completed by invoking some external physical arguments ---
in the present case the equivalence between mass dipole and
center-of-mass position.

More importantly, we find that Eq. (\ref{xikappa}) is nicely consistent with
the calculation of the ambiguity parameters by means of dimensional
regularization \cite{BDEI04,BDEI04dr}, whose results have been given in
Eq.~(\ref{resdimreg}). The dimensional regularization \textit{is} complete; it
does not need to invoke any ``external'' physical argument in order to
determine the value of all the ambiguity parameters. Nevertheless, it remains
that our result (\ref{xikappa}), based simply on a consistency argument
(within the over-all scheme) between the 3PN equations of motion on the one
hand and the 3PN radiation field on the other hand, does provide a
verification of the consistency of dimensional regularization itself.

\acknowledgments

We would like to thank the Indo-French collaboration (IFCPAR) under
which this work has been carried out.  One of us (L.B.) would like to
thank Misao Sasaki for an invitation at the Yukawa Institute for
Theoretical Physics in Kyoto University.

\bibliography{BI04mult}

\begin{thebibliography}{47}
\expandafter\ifx\csname natexlab\endcsname\relax\def\natexlab#1{#1}\fi
\expandafter\ifx\csname bibnamefont\endcsname\relax
  \def\bibnamefont#1{#1}\fi
\expandafter\ifx\csname bibfnamefont\endcsname\relax
  \def\bibfnamefont#1{#1}\fi
\expandafter\ifx\csname citenamefont\endcsname\relax
  \def\citenamefont#1{#1}\fi
\expandafter\ifx\csname url\endcsname\relax
  \def\url#1{\texttt{#1}}\fi
\expandafter\ifx\csname urlprefix\endcsname\relax\def\urlprefix{URL }\fi
\providecommand{\bibinfo}[2]{#2}
\providecommand{\eprint}[2][]{\url{#2}}

\bibitem[{\citenamefont{Blanchet et~al.}(2002)\citenamefont{Blanchet, Iyer, and
  Joguet}}]{BIJ02}
\bibinfo{author}{\bibfnamefont{L.}~\bibnamefont{Blanchet}},
  \bibinfo{author}{\bibfnamefont{B.~R.} \bibnamefont{Iyer}}, \bibnamefont{and}
  \bibinfo{author}{\bibfnamefont{B.}~\bibnamefont{Joguet}},
  \bibinfo{journal}{Phys. Rev. D} \textbf{\bibinfo{volume}{65}},
  \bibinfo{pages}{064005} (\bibinfo{year}{2002}), \eprint{gr-qc/0105098}.

\bibitem[{\citenamefont{Kalogera et~al.}(2004)\citenamefont{Kalogera, Kim,
  Lorimer, Burgay, D'Amico, Possenti, Manchester, Lyne, Joshi, McLaughlin
  et~al.}}]{kalogera}
\bibinfo{author}{\bibfnamefont{V.}~\bibnamefont{Kalogera}},
  \bibinfo{author}{\bibfnamefont{C.}~\bibnamefont{Kim}},
  \bibinfo{author}{\bibfnamefont{D.}~\bibnamefont{Lorimer}},
  \bibinfo{author}{\bibfnamefont{M.}~\bibnamefont{Burgay}},
  \bibinfo{author}{\bibfnamefont{N.}~\bibnamefont{D'Amico}},
  \bibinfo{author}{\bibfnamefont{A.}~\bibnamefont{Possenti}},
  \bibinfo{author}{\bibfnamefont{R.}~\bibnamefont{Manchester}},
  \bibinfo{author}{\bibfnamefont{A.}~\bibnamefont{Lyne}},
  \bibinfo{author}{\bibfnamefont{B.}~\bibnamefont{Joshi}},
  \bibinfo{author}{\bibfnamefont{M.}~\bibnamefont{McLaughlin}},
  \bibnamefont{et~al.}, \bibinfo{journal}{Astrophys. J.}
  \textbf{\bibinfo{volume}{601}}, \bibinfo{pages}{L179} (\bibinfo{year}{2004}),
  \eprint{astro-ph/0312101}.

\bibitem[{\citenamefont{Cutler et~al.}(1993)\citenamefont{Cutler, Apostolatos,
  Bildsten, Finn, Flanagan, Kennefick, Markovic, Ori, Poisson, Sussman
  et~al.}}]{3mn}
\bibinfo{author}{\bibfnamefont{C.}~\bibnamefont{Cutler}},
  \bibinfo{author}{\bibfnamefont{T.}~\bibnamefont{Apostolatos}},
  \bibinfo{author}{\bibfnamefont{L.}~\bibnamefont{Bildsten}},
  \bibinfo{author}{\bibfnamefont{L.}~\bibnamefont{Finn}},
  \bibinfo{author}{\bibfnamefont{E.}~\bibnamefont{Flanagan}},
  \bibinfo{author}{\bibfnamefont{D.}~\bibnamefont{Kennefick}},
  \bibinfo{author}{\bibfnamefont{D.}~\bibnamefont{Markovic}},
  \bibinfo{author}{\bibfnamefont{A.}~\bibnamefont{Ori}},
  \bibinfo{author}{\bibfnamefont{E.}~\bibnamefont{Poisson}},
  \bibinfo{author}{\bibfnamefont{G.}~\bibnamefont{Sussman}},
  \bibnamefont{et~al.}, \bibinfo{journal}{Phys. Rev. Lett.}
  \textbf{\bibinfo{volume}{70}}, \bibinfo{pages}{2984} (\bibinfo{year}{1993}).

\bibitem[{\citenamefont{Cutler and Flanagan}(1994)}]{CF94}
\bibinfo{author}{\bibfnamefont{C.}~\bibnamefont{Cutler}} \bibnamefont{and}
  \bibinfo{author}{\bibfnamefont{E.}~\bibnamefont{Flanagan}},
  \bibinfo{journal}{Phys. Rev. D} \textbf{\bibinfo{volume}{49}},
  \bibinfo{pages}{2658} (\bibinfo{year}{1994}).

\bibitem[{\citenamefont{Tagoshi and Nakamura}(1994)}]{TNaka94}
\bibinfo{author}{\bibfnamefont{H.}~\bibnamefont{Tagoshi}} \bibnamefont{and}
  \bibinfo{author}{\bibfnamefont{T.}~\bibnamefont{Nakamura}},
  \bibinfo{journal}{Phys. Rev. D} \textbf{\bibinfo{volume}{49}},
  \bibinfo{pages}{4016} (\bibinfo{year}{1994}).

\bibitem[{\citenamefont{Poisson}(1995)}]{P95}
\bibinfo{author}{\bibfnamefont{E.}~\bibnamefont{Poisson}},
  \bibinfo{journal}{Phys. Rev. D} \textbf{\bibinfo{volume}{52}},
  \bibinfo{pages}{5719} (\bibinfo{year}{1995}), \bibinfo{note}{erratum Phys.
  Rev. D {\bf 55}, 7980, (1997)}.

\bibitem[{\citenamefont{Damour et~al.}(1998)\citenamefont{Damour, Iyer, and
  Sathyaprakash}}]{DIS98}
\bibinfo{author}{\bibfnamefont{T.}~\bibnamefont{Damour}},
  \bibinfo{author}{\bibfnamefont{B.}~\bibnamefont{Iyer}}, \bibnamefont{and}
  \bibinfo{author}{\bibfnamefont{B.}~\bibnamefont{Sathyaprakash}},
  \bibinfo{journal}{Phys. Rev. D} \textbf{\bibinfo{volume}{57}},
  \bibinfo{pages}{885} (\bibinfo{year}{1998}).

\bibitem[{\citenamefont{Damour et~al.}(2001{\natexlab{a}})\citenamefont{Damour,
  Iyer, and Sathyaprakash}}]{DIS01}
\bibinfo{author}{\bibfnamefont{T.}~\bibnamefont{Damour}},
  \bibinfo{author}{\bibfnamefont{B.~R.} \bibnamefont{Iyer}}, \bibnamefont{and}
  \bibinfo{author}{\bibfnamefont{B.~S.} \bibnamefont{Sathyaprakash}},
  \bibinfo{journal}{Phys. Rev. D} \textbf{\bibinfo{volume}{63}},
  \bibinfo{pages}{044023} (\bibinfo{year}{2001}{\natexlab{a}}),
  \eprint{gr-qc/0010009}.

\bibitem[{\citenamefont{Damour et~al.}(2003)\citenamefont{Damour, Iyer,
  Jaranowski, and Sathyaprakash}}]{DIJS03}
\bibinfo{author}{\bibfnamefont{T.}~\bibnamefont{Damour}},
  \bibinfo{author}{\bibfnamefont{B.~R.} \bibnamefont{Iyer}},
  \bibinfo{author}{\bibfnamefont{P.}~\bibnamefont{Jaranowski}},
  \bibnamefont{and} \bibinfo{author}{\bibfnamefont{B.~S.}
  \bibnamefont{Sathyaprakash}}, \bibinfo{journal}{Phys. Rev. D}
  \textbf{\bibinfo{volume}{67}}, \bibinfo{pages}{064028}
  (\bibinfo{year}{2003}), \eprint{gr-qc/0211041}.

\bibitem[{\citenamefont{Buonanno
  et~al.}(2003{\natexlab{a}})\citenamefont{Buonanno, Chen, and
  Vallisneri}}]{BCV03a}
\bibinfo{author}{\bibfnamefont{A.}~\bibnamefont{Buonanno}},
  \bibinfo{author}{\bibfnamefont{Y.}~\bibnamefont{Chen}}, \bibnamefont{and}
  \bibinfo{author}{\bibfnamefont{M.}~\bibnamefont{Vallisneri}},
  \bibinfo{journal}{Phys. Rev. D} \textbf{\bibinfo{volume}{67}},
  \bibinfo{pages}{024016} (\bibinfo{year}{2003}{\natexlab{a}}),
  \eprint{gr-qc/0205122}.

\bibitem[{\citenamefont{Buonanno
  et~al.}(2003{\natexlab{b}})\citenamefont{Buonanno, Chen, and
  Vallisneri}}]{BCV03b}
\bibinfo{author}{\bibfnamefont{A.}~\bibnamefont{Buonanno}},
  \bibinfo{author}{\bibfnamefont{Y.}~\bibnamefont{Chen}}, \bibnamefont{and}
  \bibinfo{author}{\bibfnamefont{M.}~\bibnamefont{Vallisneri}},
  \bibinfo{journal}{Phys. Rev. D} \textbf{\bibinfo{volume}{67}},
  \bibinfo{pages}{104025} (\bibinfo{year}{2003}{\natexlab{b}}),
  \eprint{gr-qc/0211087}.

\bibitem[{\citenamefont{Damour}(1983)}]{D83houches}
\bibinfo{author}{\bibfnamefont{T.}~\bibnamefont{Damour}}, in
  \emph{\bibinfo{booktitle}{Gravitational Radiation}}, edited by
  \bibinfo{editor}{\bibfnamefont{N.}~\bibnamefont{Deruelle}} \bibnamefont{and}
  \bibinfo{editor}{\bibfnamefont{T.}~\bibnamefont{Piran}}
  (\bibinfo{publisher}{North-Holland Company}, \bibinfo{address}{Amsterdam},
  \bibinfo{year}{1983}), pp. \bibinfo{pages}{59--144}.

\bibitem[{\citenamefont{Hadamard}(1932)}]{Hadamard}
\bibinfo{author}{\bibfnamefont{J.}~\bibnamefont{Hadamard}},
  \emph{\bibinfo{title}{Le probl\`eme de Cauchy et les \'equations aux
  d\'eriv\'ees partielles lin\'eaires hyperboliques}}
  (\bibinfo{publisher}{Hermann}, \bibinfo{address}{Paris},
  \bibinfo{year}{1932}).

\bibitem[{\citenamefont{Schwartz}(1978)}]{Schwartz}
\bibinfo{author}{\bibfnamefont{L.}~\bibnamefont{Schwartz}},
  \emph{\bibinfo{title}{Th\'eorie des distributions}}
  (\bibinfo{publisher}{Hermann}, \bibinfo{address}{Paris},
  \bibinfo{year}{1978}).

\bibitem[{\citenamefont{Sellier}(1994)}]{Sellier}
\bibinfo{author}{\bibfnamefont{A.}~\bibnamefont{Sellier}},
  \bibinfo{journal}{Proc. R. Soc. London, Ser. A}
  \textbf{\bibinfo{volume}{445}}, \bibinfo{pages}{69} (\bibinfo{year}{1994}).

\bibitem[{\citenamefont{Sasaki}(1994)}]{Sasa94}
\bibinfo{author}{\bibfnamefont{M.}~\bibnamefont{Sasaki}},
  \bibinfo{journal}{Prog. Theor. Phys.} \textbf{\bibinfo{volume}{92}},
  \bibinfo{pages}{17} (\bibinfo{year}{1994}).

\bibitem[{\citenamefont{Tanaka et~al.}(1996)\citenamefont{Tanaka, Tagoshi, and
  Sasaki}}]{TTS96}
\bibinfo{author}{\bibfnamefont{T.}~\bibnamefont{Tanaka}},
  \bibinfo{author}{\bibfnamefont{H.}~\bibnamefont{Tagoshi}}, \bibnamefont{and}
  \bibinfo{author}{\bibfnamefont{M.}~\bibnamefont{Sasaki}},
  \bibinfo{journal}{Prog. Theor. Phys.} \textbf{\bibinfo{volume}{96}},
  \bibinfo{pages}{1087} (\bibinfo{year}{1996}).

\bibitem[{\citenamefont{Tagoshi and Sasaki}(1994)}]{TSasa94}
\bibinfo{author}{\bibfnamefont{H.}~\bibnamefont{Tagoshi}} \bibnamefont{and}
  \bibinfo{author}{\bibfnamefont{M.}~\bibnamefont{Sasaki}},
  \bibinfo{journal}{Prog. Theor. Phys.} \textbf{\bibinfo{volume}{92}},
  \bibinfo{pages}{745} (\bibinfo{year}{1994}).

\bibitem[{\citenamefont{Jaranowski and Sch\"afer}(1998)}]{JaraS98}
\bibinfo{author}{\bibfnamefont{P.}~\bibnamefont{Jaranowski}} \bibnamefont{and}
  \bibinfo{author}{\bibfnamefont{G.}~\bibnamefont{Sch\"afer}},
  \bibinfo{journal}{Phys. Rev. D} \textbf{\bibinfo{volume}{57}},
  \bibinfo{pages}{7274} (\bibinfo{year}{1998}).

\bibitem[{\citenamefont{Jaranowski and Sch\"afer}(1999)}]{JaraS99}
\bibinfo{author}{\bibfnamefont{P.}~\bibnamefont{Jaranowski}} \bibnamefont{and}
  \bibinfo{author}{\bibfnamefont{G.}~\bibnamefont{Sch\"afer}},
  \bibinfo{journal}{Phys. Rev. D} \textbf{\bibinfo{volume}{60}},
  \bibinfo{pages}{124003} (\bibinfo{year}{1999}).

\bibitem[{\citenamefont{Blanchet and Faye}(2000{\natexlab{a}})}]{BF00}
\bibinfo{author}{\bibfnamefont{L.}~\bibnamefont{Blanchet}} \bibnamefont{and}
  \bibinfo{author}{\bibfnamefont{G.}~\bibnamefont{Faye}},
  \bibinfo{journal}{Phys. Lett. A} \textbf{\bibinfo{volume}{271}},
  \bibinfo{pages}{58} (\bibinfo{year}{2000}{\natexlab{a}}),
  \eprint{gr-qc/0004009}.

\bibitem[{\citenamefont{Blanchet and Faye}(2001{\natexlab{a}})}]{BFeom}
\bibinfo{author}{\bibfnamefont{L.}~\bibnamefont{Blanchet}} \bibnamefont{and}
  \bibinfo{author}{\bibfnamefont{G.}~\bibnamefont{Faye}},
  \bibinfo{journal}{Phys. Rev. D} \textbf{\bibinfo{volume}{63}},
  \bibinfo{pages}{062005} (\bibinfo{year}{2001}{\natexlab{a}}),
  \eprint{gr-qc/0007051}.

\bibitem[{\citenamefont{Blanchet and Faye}(2000{\natexlab{b}})}]{BFreg}
\bibinfo{author}{\bibfnamefont{L.}~\bibnamefont{Blanchet}} \bibnamefont{and}
  \bibinfo{author}{\bibfnamefont{G.}~\bibnamefont{Faye}}, \bibinfo{journal}{J.
  Math. Phys.} \textbf{\bibinfo{volume}{41}}, \bibinfo{pages}{7675}
  (\bibinfo{year}{2000}{\natexlab{b}}), \eprint{gr-qc/0004008}.

\bibitem[{\citenamefont{Blanchet and Faye}(2001{\natexlab{b}})}]{BFregM}
\bibinfo{author}{\bibfnamefont{L.}~\bibnamefont{Blanchet}} \bibnamefont{and}
  \bibinfo{author}{\bibfnamefont{G.}~\bibnamefont{Faye}}, \bibinfo{journal}{J.
  Math. Phys.} \textbf{\bibinfo{volume}{42}}, \bibinfo{pages}{4391}
  (\bibinfo{year}{2001}{\natexlab{b}}), \eprint{gr-qc/0006100}.

\bibitem[{\citenamefont{Damour et~al.}(2000)\citenamefont{Damour, Jaranowski,
  and Sch\"afer}}]{DJSpoinc}
\bibinfo{author}{\bibfnamefont{T.}~\bibnamefont{Damour}},
  \bibinfo{author}{\bibfnamefont{P.}~\bibnamefont{Jaranowski}},
  \bibnamefont{and}
  \bibinfo{author}{\bibfnamefont{G.}~\bibnamefont{Sch\"afer}},
  \bibinfo{journal}{Phys. Rev. D} \textbf{\bibinfo{volume}{62}},
  \bibinfo{pages}{021501R} (\bibinfo{year}{2000}), \bibinfo{note}{erratum Phys.
  Rev. D {\bf 63}, 029903, (2001)}.

\bibitem[{\citenamefont{Damour et~al.}(2001{\natexlab{b}})\citenamefont{Damour,
  Jaranowski, and Sch\"afer}}]{DJSdim}
\bibinfo{author}{\bibfnamefont{T.}~\bibnamefont{Damour}},
  \bibinfo{author}{\bibfnamefont{P.}~\bibnamefont{Jaranowski}},
  \bibnamefont{and}
  \bibinfo{author}{\bibfnamefont{G.}~\bibnamefont{Sch\"afer}},
  \bibinfo{journal}{Phys. Lett. B} \textbf{\bibinfo{volume}{513}},
  \bibinfo{pages}{147} (\bibinfo{year}{2001}{\natexlab{b}}).

\bibitem[{\citenamefont{Blanchet
  et~al.}(2004{\natexlab{a}})\citenamefont{Blanchet, Damour, and
  Esposito-Far{\`e}se}}]{BDE04}
\bibinfo{author}{\bibfnamefont{L.}~\bibnamefont{Blanchet}},
  \bibinfo{author}{\bibfnamefont{T.}~\bibnamefont{Damour}}, \bibnamefont{and}
  \bibinfo{author}{\bibfnamefont{G.}~\bibnamefont{Esposito-Far{\`e}se}},
  \bibinfo{journal}{Phys. Rev. D} \textbf{\bibinfo{volume}{69}},
  \bibinfo{pages}{124007} (\bibinfo{year}{2004}{\natexlab{a}}),
  \eprint{gr-qc/0311052}.

\bibitem[{\citenamefont{Itoh et~al.}(2000)\citenamefont{Itoh, Futamase, and
  Asada}}]{IFA00}
\bibinfo{author}{\bibfnamefont{Y.}~\bibnamefont{Itoh}},
  \bibinfo{author}{\bibfnamefont{T.}~\bibnamefont{Futamase}}, \bibnamefont{and}
  \bibinfo{author}{\bibfnamefont{H.}~\bibnamefont{Asada}},
  \bibinfo{journal}{Phys. Rev. D} \textbf{\bibinfo{volume}{62}},
  \bibinfo{pages}{064002} (\bibinfo{year}{2000}).

\bibitem[{\citenamefont{Itoh et~al.}(2001)\citenamefont{Itoh, Futamase, and
  Asada}}]{IFA01}
\bibinfo{author}{\bibfnamefont{Y.}~\bibnamefont{Itoh}},
  \bibinfo{author}{\bibfnamefont{T.}~\bibnamefont{Futamase}}, \bibnamefont{and}
  \bibinfo{author}{\bibfnamefont{H.}~\bibnamefont{Asada}},
  \bibinfo{journal}{Phys. Rev. D} \textbf{\bibinfo{volume}{63}},
  \bibinfo{pages}{064038} (\bibinfo{year}{2001}).

\bibitem[{\citenamefont{Itoh and Futamase}(2003)}]{itoh1}
\bibinfo{author}{\bibfnamefont{Y.}~\bibnamefont{Itoh}} \bibnamefont{and}
  \bibinfo{author}{\bibfnamefont{T.}~\bibnamefont{Futamase}},
  \bibinfo{journal}{Phys. Rev. D} \textbf{\bibinfo{volume}{68}},
  \bibinfo{pages}{121501} (\bibinfo{year}{2003}).

\bibitem[{\citenamefont{Itoh}(2004)}]{itoh2}
\bibinfo{author}{\bibfnamefont{Y.}~\bibnamefont{Itoh}}, \bibinfo{journal}{Phys.
  Rev. D} \textbf{\bibinfo{volume}{69}}, \bibinfo{pages}{064018}
  (\bibinfo{year}{2004}).

\bibitem[{\citenamefont{Blanchet
  et~al.}(2004{\natexlab{b}})\citenamefont{Blanchet, Damour,
  Esposito-Far{\`e}se, and Iyer}}]{BDEI04}
\bibinfo{author}{\bibfnamefont{L.}~\bibnamefont{Blanchet}},
  \bibinfo{author}{\bibfnamefont{T.}~\bibnamefont{Damour}},
  \bibinfo{author}{\bibfnamefont{G.}~\bibnamefont{Esposito-Far{\`e}se}},
  \bibnamefont{and} \bibinfo{author}{\bibfnamefont{B.~R.} \bibnamefont{Iyer}},
  \bibinfo{journal}{Phys. Rev. Lett.} \textbf{\bibinfo{volume}{93}},
  \bibinfo{pages}{091101} (\bibinfo{year}{2004}{\natexlab{b}}),
  \eprint{gr-qc/0406012}.

\bibitem[{\citenamefont{Blanchet
  et~al.}(2004{\natexlab{c}})\citenamefont{Blanchet, Damour,
  Esposito-Far{\`e}se, and Iyer}}]{BDEI04dr}
\bibinfo{author}{\bibfnamefont{L.}~\bibnamefont{Blanchet}},
  \bibinfo{author}{\bibfnamefont{T.}~\bibnamefont{Damour}},
  \bibinfo{author}{\bibfnamefont{G.}~\bibnamefont{Esposito-Far{\`e}se}},
  \bibnamefont{and} \bibinfo{author}{\bibfnamefont{B.~R.} \bibnamefont{Iyer}}
  (\bibinfo{year}{2004}{\natexlab{c}}), \bibinfo{note}{work in preparation}.

\bibitem[{\citenamefont{Blanchet
  et~al.}(2004{\natexlab{d}})\citenamefont{Blanchet, Damour, and
  Iyer}}]{BDI04zeta}
\bibinfo{author}{\bibfnamefont{L.}~\bibnamefont{Blanchet}},
  \bibinfo{author}{\bibfnamefont{T.}~\bibnamefont{Damour}}, \bibnamefont{and}
  \bibinfo{author}{\bibfnamefont{B.~R.} \bibnamefont{Iyer}}
  (\bibinfo{year}{2004}{\natexlab{d}}), \bibinfo{note}{submitted to Class.
  Quantum Grav.}, \eprint{gr-qc/0410021}.

\bibitem[{\citenamefont{de~Andrade et~al.}(2001)\citenamefont{de~Andrade,
  Blanchet, and Faye}}]{ABF01}
\bibinfo{author}{\bibfnamefont{V.}~\bibnamefont{de~Andrade}},
  \bibinfo{author}{\bibfnamefont{L.}~\bibnamefont{Blanchet}}, \bibnamefont{and}
  \bibinfo{author}{\bibfnamefont{G.}~\bibnamefont{Faye}},
  \bibinfo{journal}{Class. Quantum Grav.} \textbf{\bibinfo{volume}{18}},
  \bibinfo{pages}{753} (\bibinfo{year}{2001}).

\bibitem[{\citenamefont{Damour et~al.}(2001{\natexlab{c}})\citenamefont{Damour,
  Jaranowski, and Sch\"afer}}]{DJSequiv}
\bibinfo{author}{\bibfnamefont{T.}~\bibnamefont{Damour}},
  \bibinfo{author}{\bibfnamefont{P.}~\bibnamefont{Jaranowski}},
  \bibnamefont{and}
  \bibinfo{author}{\bibfnamefont{G.}~\bibnamefont{Sch\"afer}},
  \bibinfo{journal}{Phys. Rev. D} \textbf{\bibinfo{volume}{63}},
  \bibinfo{pages}{044021} (\bibinfo{year}{2001}{\natexlab{c}}).

\bibitem[{\citenamefont{Blanchet}(1995)}]{B95}
\bibinfo{author}{\bibfnamefont{L.}~\bibnamefont{Blanchet}},
  \bibinfo{journal}{Phys. Rev. D} \textbf{\bibinfo{volume}{51}},
  \bibinfo{pages}{2559} (\bibinfo{year}{1995}), \eprint{gr-qc/9501030}.

\bibitem[{\citenamefont{Blanchet}(1998{\natexlab{a}})}]{B98mult}
\bibinfo{author}{\bibfnamefont{L.}~\bibnamefont{Blanchet}},
  \bibinfo{journal}{Class. Quant. Grav.} \textbf{\bibinfo{volume}{15}},
  \bibinfo{pages}{1971} (\bibinfo{year}{1998}{\natexlab{a}}),
  \eprint{gr-qc/9801101}.

\bibitem[{\citenamefont{Poujade and Blanchet}(2002)}]{PB02}
\bibinfo{author}{\bibfnamefont{O.}~\bibnamefont{Poujade}} \bibnamefont{and}
  \bibinfo{author}{\bibfnamefont{L.}~\bibnamefont{Blanchet}},
  \bibinfo{journal}{Phys. Rev. D} \textbf{\bibinfo{volume}{65}},
  \bibinfo{pages}{124020} (\bibinfo{year}{2002}), \eprint{gr-qc/0112057}.

\bibitem[{\citenamefont{Blanchet and Damour}(1986)}]{BD86}
\bibinfo{author}{\bibfnamefont{L.}~\bibnamefont{Blanchet}} \bibnamefont{and}
  \bibinfo{author}{\bibfnamefont{T.}~\bibnamefont{Damour}},
  \bibinfo{journal}{Phil. Trans. Roy. Soc. Lond. A}
  \textbf{\bibinfo{volume}{320}}, \bibinfo{pages}{379} (\bibinfo{year}{1986}).

\bibitem[{\citenamefont{Blanchet}(1998{\natexlab{b}})}]{B98tail}
\bibinfo{author}{\bibfnamefont{L.}~\bibnamefont{Blanchet}},
  \bibinfo{journal}{Class. Quant. Grav.} \textbf{\bibinfo{volume}{15}},
  \bibinfo{pages}{113} (\bibinfo{year}{1998}{\natexlab{b}}),
  \eprint{gr-qc/9710038}.

\bibitem[{\citenamefont{Blanchet and Damour}(1989)}]{BD89}
\bibinfo{author}{\bibfnamefont{L.}~\bibnamefont{Blanchet}} \bibnamefont{and}
  \bibinfo{author}{\bibfnamefont{T.}~\bibnamefont{Damour}},
  \bibinfo{journal}{Annales Inst. H. Poincar\'e Phys. Th\'eor.}
  \textbf{\bibinfo{volume}{50}}, \bibinfo{pages}{377} (\bibinfo{year}{1989}).

\bibitem[{\citenamefont{Damour and Iyer}(1991)}]{DI91b}
\bibinfo{author}{\bibfnamefont{T.}~\bibnamefont{Damour}} \bibnamefont{and}
  \bibinfo{author}{\bibfnamefont{B.~R.} \bibnamefont{Iyer}},
  \bibinfo{journal}{Phys. Rev. D} \textbf{\bibinfo{volume}{43}},
  \bibinfo{pages}{3259} (\bibinfo{year}{1991}).

\bibitem[{\citenamefont{Blanchet et~al.}(1998)\citenamefont{Blanchet, Faye, and
  Ponsot}}]{BFP98}
\bibinfo{author}{\bibfnamefont{L.}~\bibnamefont{Blanchet}},
  \bibinfo{author}{\bibfnamefont{G.}~\bibnamefont{Faye}}, \bibnamefont{and}
  \bibinfo{author}{\bibfnamefont{B.}~\bibnamefont{Ponsot}},
  \bibinfo{journal}{Phys. Rev. D} \textbf{\bibinfo{volume}{58}},
  \bibinfo{pages}{124002} (\bibinfo{year}{1998}), \eprint{gr-qc/9804079}.

\bibitem[{\citenamefont{Gel'fand and Shilov}(1964)}]{gelfand}
\bibinfo{author}{\bibfnamefont{I.~M.} \bibnamefont{Gel'fand}} \bibnamefont{and}
  \bibinfo{author}{\bibfnamefont{G.~E.} \bibnamefont{Shilov}},
  \emph{\bibinfo{title}{Generalized functions}} (\bibinfo{publisher}{Academic
  Press}, \bibinfo{address}{New York}, \bibinfo{year}{1964}).

\bibitem[{\citenamefont{Gradshteyn and Ryzhik}(1980)}]{GZ}
\bibinfo{author}{\bibfnamefont{I.}~\bibnamefont{Gradshteyn}} \bibnamefont{and}
  \bibinfo{author}{\bibfnamefont{I.}~\bibnamefont{Ryzhik}},
  \emph{\bibinfo{title}{Table of Integrals, Series and Products}}
  (\bibinfo{publisher}{Academic Press}, \bibinfo{year}{1980}).

\bibitem[{\citenamefont{Blanchet and Iyer}(2003)}]{BI03CM}
\bibinfo{author}{\bibfnamefont{L.}~\bibnamefont{Blanchet}} \bibnamefont{and}
  \bibinfo{author}{\bibfnamefont{B.~R.} \bibnamefont{Iyer}},
  \bibinfo{journal}{Class. Quant. Grav.} \textbf{\bibinfo{volume}{20}},
  \bibinfo{pages}{755} (\bibinfo{year}{2003}), \eprint{gr-qc/0209089}.

\end{thebibliography}

\end{document}